\documentclass[reprint, amssymb, amsmath, pre, showkeys]{revtex4-1}
\usepackage{graphicx, color, hyperref, array}

\newcommand{\beq}{\begin{equation}}
\newcommand{\eeq}{\end{equation}}
\newcommand{\sat}{\mathrm{sat}}

\newcommand{\Ybar}{\bar{Y}}
\newcommand{\tbar}{\bar{\tau}}
\newcommand{\new}[1]{{#1}}

\begin{document}

\title{Growth tradeoffs produce complex microbial communities \new{on a single limiting resource}}


\author{Michael Manhart}
\email[To whom correspondence should be addressed.  Email: ]{mmanhart@fas.harvard.edu}
\affiliation{Department of Chemistry and Chemical Biology, Harvard University, Cambridge, MA 02138, USA}
\author{Eugene I. Shakhnovich}
\affiliation{Department of Chemistry and Chemical Biology, Harvard University, Cambridge, MA 02138, USA}
\date{\today}


\begin{abstract}
\new{The relationship between the dynamics of a community and its constituent pairwise interactions is a fundamental problem in ecology.  
Higher-order ecological effects beyond pairwise interactions may be key to complex ecosystems, but mechanisms to produce these effects remain poorly understood.  Here we show that higher-order effects can arise from variation in multiple microbial growth traits, such as lag times and growth rates, on a single limiting resource with no other interactions.}
These effects 
produce a range of ecological phenomena: an unlimited number of strains can \new{exhibit multistability and neutral coexistence}, potentially with a single keystone strain; strains that coexist in pairs do not coexist all together; and the champion of all pairwise competitions may not dominate in a mixed community.  
Since variation in multiple growth traits is ubiquitous in microbial populations due to pleiotropy and non-genetic variation, our results indicate these higher-order effects may also be widespread, especially in laboratory ecology and evolution experiments.  
\end{abstract}

\maketitle



	
	\new{Complex communities of many distinct species or strains abound in both nature~\cite{HMPC2012, Sunagawa2015} and the laboratory~\cite{Frentz2015, Good2017}.  One of the most fundamental problems in ecology is to understand the relationship between a community's behavior and the pairwise interactions of its constituents~\cite{Faust2012, Bucci2014, Momeni2017, Friedman2017}.  In particular, the key question is whether these pairwise interactions largely determine the behavior of the community as a whole.  However, ecologists have long considered the possibility of ``higher-order'' effects such that the interaction between pairs of strains can be altered by the presence of additional strains~\cite{Billick1994, Wootten1994, Momeni2017, Friedman2017, Mayfield2017}.  These higher-order effects may cause a community to be fundamentally different than the sum of its pairwise interactions and can play an important role in stabilizing coexisting communities~\cite{Bairey2016, Grilli2017}.  Although these higher-order effects may be essential to accurately predict the ecological and evolutionary dynamics of a population, they remain poorly characterized in general.}

     The relative simplicity and experimental tractability of microbes make them convenient for studying this problem.  Most well-known ecological effects in microbes are mediated by cross-feeding interactions or the consumption of multiple resources~\cite{Widder2016}.  
For example, long-term coexistence of distinct strains is often believed to depend on the existence of at least as many resource types as coexisting strains, according to the ``principle of competitive exclusion''~\cite{Hardin1960, Levin1970}.  However, theoretical and experimental work has demonstrated that tradeoffs in life-history traits alone --- for example, growing quickly at low concentration of a resource versus growing quickly at high concentrations, but with only a single resource type and no other interactions --- are sufficient to produce not only stable coexistence of two strains~\cite{Levin1972, Stewart1973, Turner1996, Smith2011} but also non-transitive selection~\cite{Manhart2018}, in which pairwise competitions of strains can exhibit ``rock-paper-scissors'' behaviors~\cite{Verhoef2010}.  

     Variation in multiple growth traits, such as lag time, exponential growth rate, and yield (resource efficiency), is pervasive in microbial populations~\cite{Novak2006, Warringer2011, Jasmin2012a}.  Not only are single mutations known to be pleiotropic with respect to these traits~\cite{Fitzsimmons2010, Adkar2017}, but even genetically-identical lineages may demonstrate significant variation~\cite{LevinReisman2010, Ziv2013}.  The ecological effects of such variation, however, are unknown in large populations with many distinct strains simultaneously competing, as is generally the case for microbes.  
     
     Here we study a model that shows how covariation in growth traits can produce complex microbial communities without any interactions among cells beyond competition for a single limiting resource.  We focus on variation in lag times, exponential growth rates, and yields since they are the traits most easily measured by growth curves of individual strains~\cite{Zwietering1990}.  We show that covariation in these traits creates higher-order effects such that the magnitude and even the sign of the selection coefficient between a pair of strains may be changed by the presence of a third strain. 
These higher-order effects can produce nontrivial ecological phenomena: an unlimited number of strains can \new{form a multistable community or neutrally} coexist, potentially with a single ``keystone'' strain stabilizing the community~\cite{Power1996, Fisher2014};  
strains that coexist in pairs do not coexist in a community all together; and the champion of all pairwise competitions may not dominate in a mixed community.  Our model can be combined with high-throughput measurements of microbial growth traits to make more accurate predictions of the distribution of ecological effects and, in turn, evolutionary dynamics.  Altogether these results show how fundamental properties of microbial growth are sufficient to generate complex ecological behavior, underscoring the necessity of considering ecology in studies of microbial evolution.


\section*{Results}


\subsection*{Minimal model of microbial growth and competition over serial dilutions}

\begin{figure}
\centering\includegraphics[width=\columnwidth]{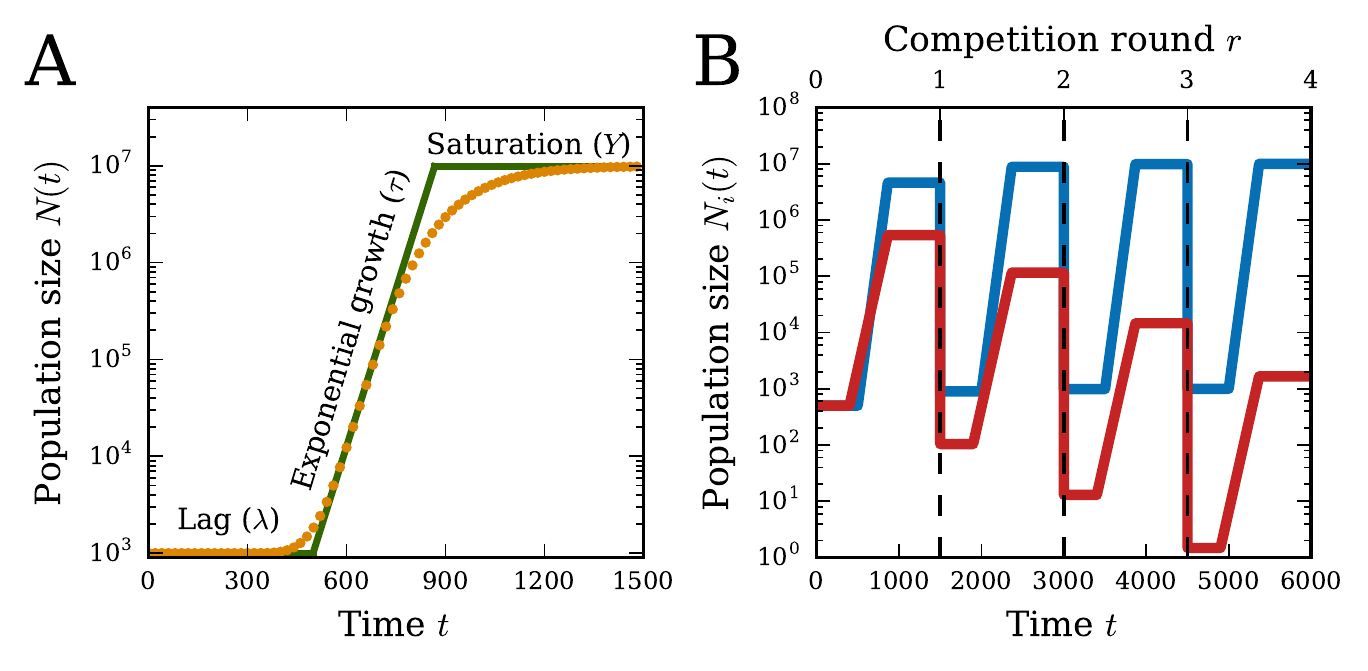}
\caption{
\textbf{Model of growth and selection.}  
(A)~Approximation of a hypothetical growth curve (orange points) by the minimal three-phase model (green; see \emph{Methods}).  Each phase is characterized by a quantitative trait: lag time $\lambda$, growth time $\tau$ (reciprocal growth rate), and yield $Y$ at saturation.  
(B)~Growth curves of two competing strains over multiple rounds of competition in the model.  Vertical dashed lines mark the beginning of each round, where the population is diluted down to the same initial population size with new resources (\emph{Methods}).  
}
\label{fig:growth_model}
\end{figure}

     We consider a microbial population consisting of multiple strains with distinct growth traits, all competing for a single limiting resource.  These strains may represent different microbial species, mutants of the same species, or even genetically-identical cells with purely phenotypic variation.  We approximate the growth of each strain $i$ by the minimal model in Fig.~\ref{fig:growth_model}A, defined by a lag time $\lambda_i$, exponential growth time $\tau_i$ (reciprocal growth rate, or time for the strain to grow $e$-fold), and yield $Y_i$, which is the population size supported per unit resource (\emph{Methods})~\cite{Vasi1994}.  \new{We assume resources are consumed in proportion to the total number of cells; it is straightforward to modify the model to other modes of resource consumption, such as where resources are consumed per cell division~\cite{Manhart2018}.}  Therefore the amount of resources strain $i$ has consumed by time $t$ is $N_i(t)/Y_i$, where $N_i(t)$ is the population size of strain $i$.  Growth stops when the amount of resources consumed by all strains equals the initial amount of resources; we define the initial density of resources per cell as $\rho$.  Although it is possible to consider additional growth traits such as a death rate or consumption of a secondary resource, here we focus on the minimal set of growth traits $\lambda_i$, $\tau_i$, and $Y_i$ since they are most often reported in microbial phenotyping experiments~\cite{Warringer2011, Zackrisson2016}.  See Table~\ref{table:notation} for a summary of all key notation.
     
     The selection coefficient between a pair of strains $i$ and $j$ measures their relative ability to compete for resources~\cite{Crow1970, Chevin2011}:
 
\beq
s_{ij} = \log\bigg(\frac{x_i'}{x_j'}\bigg) - \log\bigg(\frac{x_i}{x_j}\bigg),
\label{eq:s_def}
\eeq

\noindent where $x_i$ is the density (\new{dimensionless fraction of the whole population}) of strain $i$ at the beginning of the competition and $x_i'$ is the density at the end.  If new resources periodically become available, as occur in both laboratory evolution experiments and natural ``seasonal'' environments~\cite{Vasi1994}, then the population will undergo cycles of lag, growth, and saturation (Fig.~\ref{fig:growth_model}B).  Each round of competition begins with the same initial density of resources $\rho$.  The population grows until all the resources are consumed, and then it is diluted down to the original size again; \new{we assume the time to resource depletion is always shorter than the time between dilutions.}  We also assume the growth traits $\lambda_i$, $\tau_i$, and $Y_i$ of each strain remain the same over multiple competition rounds.  The selection coefficients in Eq.~\ref{eq:s_def} measure the rate of change of a strain's density $x_i$ over many rounds of these competitions (\emph{Methods}).

\begin{table}
\begin{tabular}{|>{\raggedright\arraybackslash\hspace{0pt}}m{0.6\columnwidth}|>{\raggedright\arraybackslash\hspace{0pt}}m{0.3\columnwidth}|}
\hline&\\[-1em]
\textbf{Definition} & \textbf{Notation} \\
\hline&\\[-1em]
Lag time of strain $i$ & $\lambda_i$ \\&\\[-1em]
\hline&\\[-1em]
Exponential growth time (reciprocal growth rate) of strain $i$ & $\tau_i$ \\&\\[-1em]
\hline&\\[-1em]
Yield (cells per resource) of strain $i$ & $Y_i$ \\&\\[-1em]
\hline&\\[-1em]
Density \new{(fraction of population)} of strain $i$ at beginning of competition round & $x_i$ \\&\\[-1em]
\hline&\\[-1em]
Density of resources per cell at beginning of competition round & $\rho$ \\&\\[-1em]
\hline&\\[-1em]
Effective exponential growth time of whole population (harmonic mean) & $\tbar = \frac{\sum_k \frac{x_k}{Y_k}}{\sum_k \frac{x_k}{Y_k \tau_k}}$ \\&\\[-1em]
\hline&\\[-1em]
Effective yield of whole population (harmonic mean) & $\Ybar = \frac{1}{\sum_k \frac{x_k}{Y_k}}$ \\&\\[-1em]
\hline&\\[-1em]
Growth-lag tradeoff & $c = -\left(\frac{\lambda_i - \lambda_j}{\tau_i - \tau_j}\right)$ \\
[0.5em]\hline
\end{tabular}
\caption{\textbf{Summary of key notation.}}
\label{table:notation}
\end{table}

\subsection*{Contribution of multiple growth traits to selection}

     We can solve for the selection coefficients in Eq.~\ref{eq:s_def} in terms of the strains' traits $\{\lambda_k, \tau_k, Y_k\}$, the initial strain densities $\{x_k\}$, and the initial density of resources per cell $\rho$ (\emph{Supplementary Methods} Sec.~S1): 
     
\begin{subequations}
\begin{align}
s_{ij} & \approx s_{ij}^\mathrm{lag} + s_{ij}^\mathrm{growth} + \sum_k s_{ijk}^\mathrm{coupling}, \label{eq:s_decomposition}\\
\intertext{where}
s_{ij}^\mathrm{lag} & = -\frac{\tbar}{\tau_i\tau_j} \Delta\lambda_{ij}, \nonumber\\
s_{ij}^\mathrm{growth} & = -\frac{\tbar}{\tau_i\tau_j} \Delta\tau_{ij} \log \left(\rho \Ybar\right), \label{eq:s_components}\\
s_{ijk}^\mathrm{coupling} & = -\frac{\tbar \Ybar}{\tau_i\tau_j} \frac{x_k}{\tau_k Y_k} \left(\Delta\tau_{ik} \Delta\lambda_{kj} - \Delta\lambda_{ik} \Delta\tau_{kj}\right). \nonumber
\end{align}
\label{eq:s_formula}
\end{subequations}




     
\noindent Here $\Delta\lambda_{ij} = \lambda_i - \lambda_j$ and $\Delta\tau_{ij} = \tau_i - \tau_j$ denote the pairwise differences in lag and growth times, while

\beq
\tbar = \frac{\sum_k \frac{x_k}{Y_k}}{\sum_k \frac{x_k}{\tau_k Y_k}}, \qquad \Ybar = \frac{1}{\sum_k \frac{x_k}{Y_k}}
\label{eq:tYbar}
\eeq

\noindent are, respectively, the effective exponential growth time (reciprocal growth rate) and effective yield for the whole population (\emph{Supplementary Methods} Sec.~S2).  Since both of these quantities are harmonic means over the population, they are dominated by the smallest trait values.  Therefore the effective growth time $\tbar$ for the whole population will be close to the growth time of the fastest-growing strain (smallest $\tau_k$), while the effective yield $\Ybar$ will generally be close to the yield of the least-efficient strain (smallest $Y_k$).

\begin{figure}
\centering\includegraphics[width=\columnwidth]{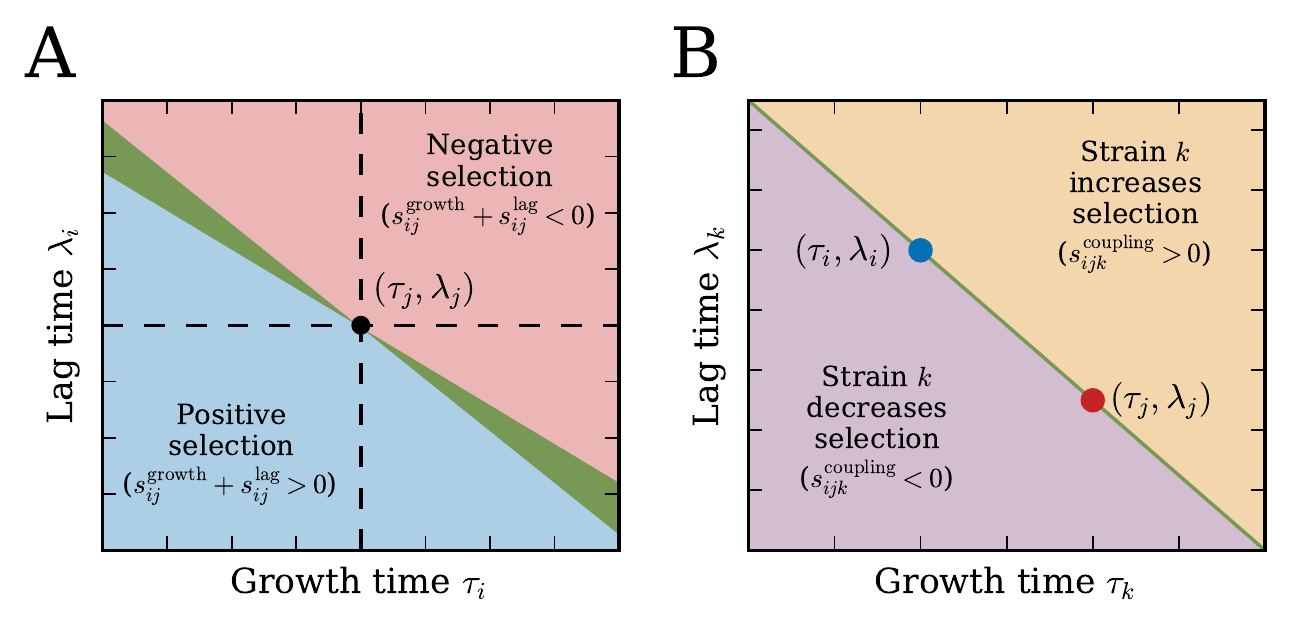}
\caption{
\textbf{Selection in growth-lag trait space.}   
(A)~Diagram of selection on growth and lag for strain $i$ relative to strain $j$.  Trait values of strain $j$ are marked by a black dot in the center.  If the traits of strain $i$ lie in the blue region, $i$ is positively selected over strain $j$, while if strain $i$ lies in the red region, it is negatively selected.  If strain $i$ lies in the green region, it is conditionally neutral with $j$ (positively selected at some densities and negatively selected at others).  
(B)~Diagram of growth and lag times for strain $k$ relative to two other strains $i$ (blue) and $j$ (red).  If the traits of strain $k$ lie in the orange region (above the straight line joining $i$ and $j$), then its coupling term $s_{ijk}^\mathrm{coupling}$ (Eq.~\ref{eq:s_components}) increases the total selection coefficient of $i$ over $j$, while if $k$ lies in the violet region (below the straight line), then it decreases the selection of $i$ over $j$.  
}
\label{fig:selection}
\end{figure}

     As Eq.~\ref{eq:s_formula} indicates, selection consists of three distinct additive components.  The first is selection on the lag phase $s_{ij}^\mathrm{lag}$, which is nonzero only if $i$ and $j$ have unequal lag times.  The second is selection on the growth phase $s_{ij}^\mathrm{growth}$, which is similarly nonzero only if $i$ and $j$ have unequal growth times.  The relative magnitude of selection on growth versus lag is modulated by the density of resources $\rho$ and the effective population yield $\Ybar$:
     
\beq
\frac{s_{ij}^\mathrm{growth}}{s_{ij}^\mathrm{lag}} = \frac{\Delta\tau_{ij}}{\Delta\lambda_{ij}} \log \left(\rho\Ybar\right).
\label{eq:relative_selection}
\eeq

\noindent In particular, increasing the resources $\rho$ leads to an increase in the magnitude of relative selection on growth versus lag, since it means the growth phase occupies a greater portion of the total competition time.  

     If $i$ and $j$ are the only two strains present, then the total selection on strain $i$ relative to $j$ is the net effect of selection on the lag and growth phases: $s_{ij} = s_{ij}^\mathrm{lag} + s_{ij}^\mathrm{growth}$~\cite{Manhart2018}.  Figure~\ref{fig:selection}A qualitatively shows this selection coefficient as a function of strain $i$'s lag and growth traits relative to those of strain $j$.  If strain $i$'s traits fall in the blue region, the overall selection on it relative to strain $j$ will be positive, while if strain $i$'s traits fall in the red region, it will be negatively selected relative to strain $j$.  Between these two regions lies a ``conditionally-neutral'' region (green), where strain $i$ will be positively selected at some densities and negatively selected at others~\cite{Manhart2018}.  The slope of the conditionally-neutral region is $\log\left(\rho\Ybar\right)$ according to Eq.~\ref{eq:relative_selection}.


\subsection*{Pairwise selection coefficients are modified by additional strains through higher-order effects}

     If more than two distinct strains are present, then selection between $i$ and $j$ is modified by higher-order effects from the other strains.  \new{These higher-order effects are separate from the effects of increasing the initial population size upon addition of more strains, which simply decreases the initial density of resources $\rho$; we therefore hold $\rho$ constant (i.e., by scaling up the total amount of resources or scaling down the initial population size for each strain) when considering the addition of another strain.}  The higher-order modifications occur through three mechanisms, all fundamentally a consequence of having a finite resource.  The first mechanism is through changes to the effective population growth time $\tbar$, which rescales all selection coefficients (Eq.~\ref{eq:s_components}).  For example, the addition of a strain with much faster growth will reduce the time all strains have to grow (Eq.~\ref{eq:tYbar}), and thereby decrease the magnitude of all selection coefficients.  The second modification is through the effective population yield $\Ybar$.  Like $\tbar$, $\Ybar$ is a harmonic mean over strains, and similarly it will be significantly reduced if a strain with very low yield is added.  This may change even the signs of some selection coefficients since changes in $\Ybar$ modify the relative selection on growth versus lag between strains (Eq.~\ref{eq:relative_selection}).
     
     Higher-order effects in $\tbar$ and $\Ybar$ are non-specific in the sense that these parameters are shared by all pairs of strains in the population.  In contrast, the third type of modification is through the terms $s_{ijk}^\mathrm{coupling}$, which couple the relative growth and lag traits of a pair $i$ and $j$ with a third strain $k$ (Eq.~\ref{eq:s_components}).  This effect is specific, since each additional strain $k$ modifies the competition between $i$ and $j$ differently, depending on its growth traits and density $x_k$.  We can interpret this effect graphically by considering the space of growth and lag times for strains $i$, $j$, and $k$ (Fig.~\ref{fig:selection}B).  If strain $k$ lies above the straight line connecting strains $i$ and $j$ in growth-lag trait space, then the coupling term will increase selection on whichever strain between $i$ and $j$ has faster growth (assumed to be strain $i$ in the figure).  This is because strain $k$ has relatively slow growth or long lag compared to $i$ and $j$, thus using fewer resources than \new{if the strains all had the same lag times or growth times}.  This then leaves more resources for $i$ and $j$, which effectively increases the selection on growth rate between the two strains beyond the $s_{ij}^\mathrm{growth}$ term.  If strain $k$ instead lies below the straight line, then it increases selection on the strain with slower growth, since $k$ uses more resources than \new{if the strains all had the same lag times or growth times}.  For example, even if strain $i$ has both better growth and better lag compared to strain $j$, a third strain $k$ could actually reduce this advantage by having sufficiently short lag.  
Note that the coupling term is zero if all three strains have equal growth times or equal lag times.  These coupling effects will furthermore be small if the relative differences in growth and lag traits are small, since $s_{ijk}^\mathrm{coupling}$ is quadratic in $\Delta\tau$ and $\Delta\lambda$ while $s_{ij}^\mathrm{lag}$ and $s_{ij}^\mathrm{growth}$ are linear.  In the following sections, we will demonstrate how these three higher-order mechanisms lead to nontrivial ecological dynamics.

\subsection*{Growth tradeoffs enable \new{multistability and coexistence} of many strains on a single resource}

\begin{figure*}
\centering\includegraphics[width=\textwidth]{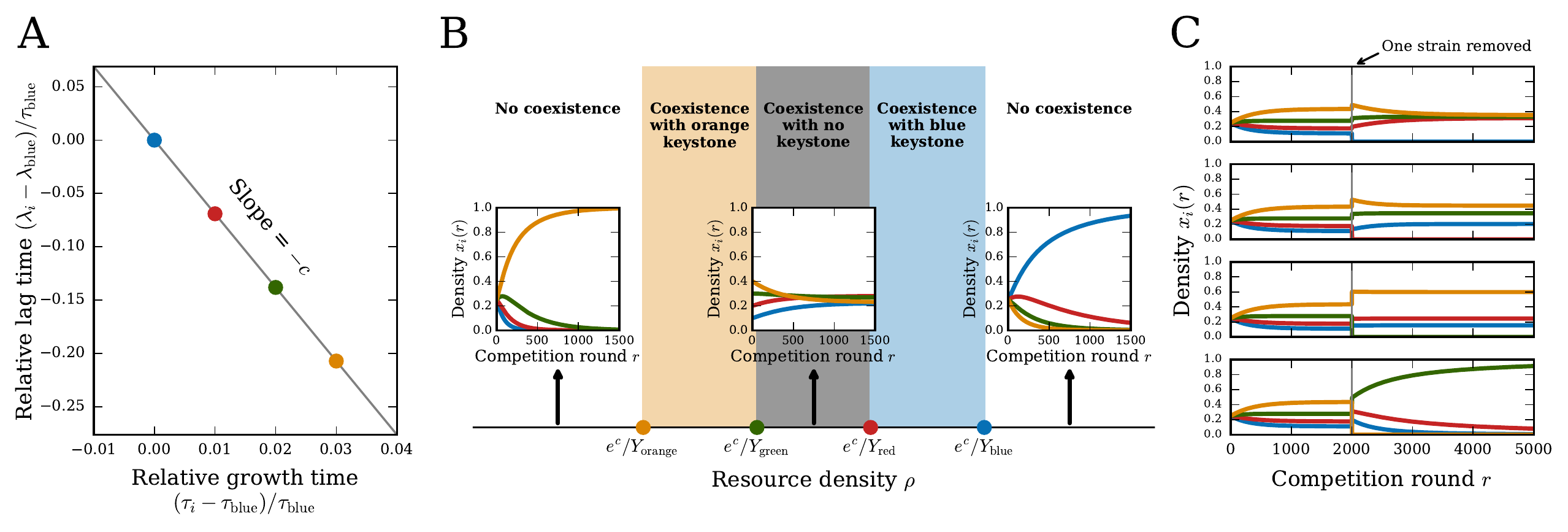}
\caption{
\textbf{Coexistence of multiple strains on a single resource.}  
(A)~Growth and lag times of four strains (blue, red, green, orange).  For them to coexist in a community, these traits must have a linear tradeoff with slope $-c$ (Eq.~\ref{eq:growth_lag_tradeoff}).  
(B)~Diagram of competition outcomes as a function of resource density $\rho$.  Each inset shows the dynamics of the strains' densities $x_i(r)$ over rounds of competition $r$ for a particular value of $\rho$.  All four strains will coexist if $\rho$ is in the range defined by Eq.~\ref{eq:rho_bounds} (shaded regions).  If $\rho$ is in the orange or blue regions, coexistence hinges on a single keystone strain of corresponding color (orange or blue), while if $\rho$ is in the gray region, coexistence is robust to loss of any single strain.  
(C)~Density dynamics of the same four strains with resource density $\rho$ in the orange region of (B), so that the orange strain is the keystone.  All four strains coexist together at first, then at competition round 2000 one strain is removed (different in each panel) and the remaining strains are allowed to reach their steady state.  
See \emph{Supplementary Methods} Sec.~S9 for parameter values.
}
\label{fig:coexistence}
\end{figure*}

     Selection is frequency-dependent since each $s_{ij}$ in Eq.~\ref{eq:s_formula} depends on the densities $\{x_k\}$~\cite{Manhart2018}.  It is therefore possible for a community of strains \new{to demonstrate multistability, including neutral coexistence} (\emph{Supplementary Methods} Sec.~S3, Fig.~S1).  \new{There can be an unlimited number of distinct strains as long as they share} a linear tradeoff between lag and growth times (Fig.~\ref{fig:coexistence}A):
     
\beq
\lambda_i = -c\tau_i + \mathrm{constant}
\label{eq:growth_lag_tradeoff}
\eeq

\noindent for all $i$ and some parameter $c > 0$, which we define as the growth-lag tradeoff.  The resource density $\rho$ must also fall in the range (Fig.~\ref{fig:coexistence}B)

\beq
\frac{e^c}{\max_k Y_k} < \rho < \frac{e^c}{\min_k Y_k}.
\label{eq:rho_bounds}
\eeq

\noindent Note that $\rho > 1/\min_k Y_k$ is necessary as well, since if $\rho$ is below this limit there will be insufficient resources for some strains to grow at all.  Since this limit is always lower than the upper bound in Eq.~\ref{eq:rho_bounds} (because $c > 0$), there will always be some range of $\rho$ at which all strains coexist.  While real strains will not exactly obey Eq.~\ref{eq:growth_lag_tradeoff}, even noisy tradeoffs cause effective coexistence over a finite time scale (\emph{Supplementary Methods} Sec.~S3, Fig.~S2).

     Intuitively, coexistence occurs because strains consume resources in such a way to exactly balance selection on lag and growth for all pairs of strains.  The linear growth-lag tradeoff across all strains from Eq.~\ref{eq:growth_lag_tradeoff} causes the higher-order coupling terms $s_{ijk}^\mathrm{coupling}$ of the selection coefficient to be zero (Fig.~\ref{fig:selection}B).  It also means there is some value of the effective yield $\Ybar$ that will enable $s_{ij}^\mathrm{growth} + s_{ij}^\mathrm{lag} = 0$ for all pairs $i$ and $j$; this critical value of the effective yield is $\Ybar = e^c/\rho$ (Eq.~\ref{eq:relative_selection}, \emph{Supplementary Methods} Sec.~S3).  The constraint on resource density $\rho$ (Eq.~\ref{eq:rho_bounds}) ensures that the population can actually achieve this required effective yield given the yield values of the individual strains.
     

     \new{If there are $M$ strains in the community satisfying the criteria in Eqs.~\ref{eq:growth_lag_tradeoff} and~\ref{eq:rho_bounds}, then there is an $(M-2)$-dimensional space of densities at which all selection coefficients are zero (\emph{Supplementary Methods} Sec.~S3, Fig.~S1).  Density perturbations within this space are therefore neutral, while perturbations orthogonal to this space will be stable if there is also a tradeoff in growth and yield (\emph{Supplementary Methods} Sec.~S4, Fig.~S3A), in addition to the growth-lag tradeoff (Eq.~\ref{eq:growth_lag_tradeoff}).  Therefore an unlimited number of strains can neutrally coexist within this space of densities until genetic drift eventually leads to extinction of all but two strains.  However, the time scale of this neutral coexistence will typically be very long compared to laboratory experiments or the times by which new mutations will arise or environmental changes will occur, since the time scale of genetic drift is of order the bottleneck population size (in units of competition rounds).  If growth and yield have a synergy across strains rather than a tradeoff, the community will be multistable, dominated by different individual strains or pairs of strains depending on the initial conditions (Fig.~S3B,C).}


\subsection*{\new{Coexistence} may hinge on a single keystone strain}

     Besides small fluctuations in densities, an even stronger perturbation to a community is to remove one strain entirely.  The stability of ecosystems in response to removal of a strain or species has long been an important problem in ecology; in particular, species whose removal leads to community collapse are known as ``keystone'' species due to their importance in stabilizing the community~\cite{Power1996, Fisher2014}.  
     
     Coexisting communities in our model will have a keystone strain for a certain range of resource density $\rho$.  Figure~\ref{fig:coexistence}B shows a diagram of competition outcomes across $\rho$ values for four hypothetical strains (blue, red, green, orange): if $\rho$ is in the orange or blue ranges, then removal of the strain of corresponding color (orange or blue) will cause rapid collapse of the community (all remaining strains but one will go extinct), since $\rho$ will no longer satisfy Eq.~\ref{eq:rho_bounds} for the remaining strains.  Therefore the orange or blue strain is the keystone.  However, if $\rho$ is within the gray region, then the community is robust to removal of any single strain.  This shows that the keystone must always be the least- or most-efficient strain (smallest or highest yield $Y_k$) in the community.  Figure~\ref{fig:coexistence}C shows the population dynamics with each strain removed from a coexisting community where the orange strain is the keystone.

     Besides removal of an existing strain, another important perturbation to a community is invasion of a new strain, either by migration or from a mutation.  
If the growth and lag times of the invader lie above the diagonal line formed by the coexisting strains' traits (e.g., as in Fig.~\ref{fig:coexistence}A), then the invader will quickly go extinct (\emph{Supplementary Methods} Sec.~S5).  This would be true even if the invader has a growth time or lag time shorter than those of all the resident strains.  On the other hand, if the invader lies below the diagonal line in growth-lag trait space, then it will either take over the population entirely or coexist with one of the resident strains if it is sufficiently close to the diagonal line.  It cannot coexist with more than one of the resident strains, since all three points by assumption will not lie on a straight line in the growth-lag trait space.


\subsection*{Pairwise competitions do not predict community behavior}

     A fundamental issue for microbial ecology and evolution is whether pairwise competitions are sufficient to predict how a whole community will behave~\cite{Billick1994, Momeni2017, Friedman2017}.  For example, if several strains coexist in pairs, will they coexist all together?  Or if a single strain dominates all pairwise competitions, will it also dominate in the mixed community?  We now show that competition for a single limiting resource with tradeoffs in growth traits is sufficient to confound these types of predictions due to the higher-order effects in the selection coefficient (Eq.~\ref{eq:s_formula}). 
     
     \textbf{Strains that coexist in pairs will generally not coexist all together.}  Strains $i$ and $j$ that coexist as a pair are characterized by a particular growth-lag tradeoff $c_{ij} = -\Delta\lambda_{ij}/\Delta\tau_{ij}$ (Eq.~\ref{eq:growth_lag_tradeoff}).  For a set of these pairs to coexist all together, these tradeoffs must all be equal, which will generally not be the case.  However, if the growth-lag tradeoffs are equal for all pairs, then the strains can indeed coexist in a community, but not at the same resource densities as for the pairs (\emph{Supplementary Methods} Sec.~S6).
     
\begin{figure}
\centering\includegraphics[width=\columnwidth]{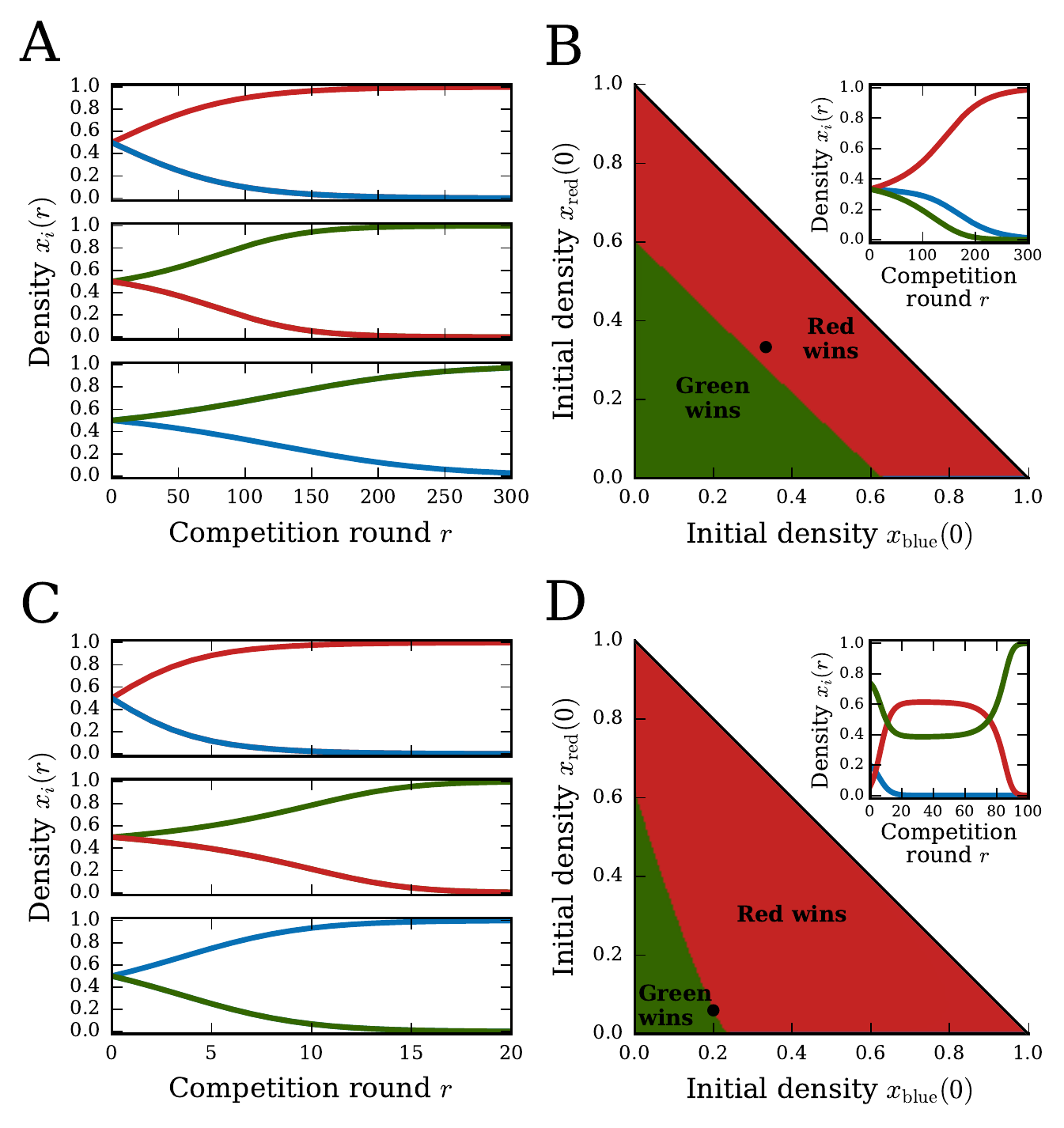}
\caption{
\textbf{Pairwise competitions do not predict community behavior.}  
(A, B)~Example of three strains (blue, red, green) with a single pairwise champion (green).  Panel (A) shows density dynamics $x_i(r)$ for binary competitions, while panel (B) shows outcome of ternary competition as a function of initial conditions: red marks space of initial densities where the red strain eventually wins, while green marks initial densities where green eventually wins.  Inset: density dynamics starting from equal initial densities (marked by black dot in main panel).  
(C, D)~Same as (A, B), but for three strains without a pairwise champion (non-transitivity).  
See \emph{Supplementary Methods} Sec.~S9 for parameter values.
}
\label{fig:binary_ternary_comparisons}
\end{figure}

     \textbf{Pairwise champion may not dominate in the community.}  In a collection of strains, there may be one ``champion'' strain that wins all pairwise competitions.  This champion, however, may not prevail in a mixed competition of all strains.  For example, in Fig.~\ref{fig:binary_ternary_comparisons}A 
the green strain beats the blue and red strains individually with a ``hoarding'' strategy --- slower growth, but shorter lag with lower yield --- but together the blue and red strains consume resources efficiently enough to use their faster growth rates to beat green (Fig.~\ref{fig:binary_ternary_comparisons}B).  This is a unique consequence of higher-order effects \new{in the selection coefficients with pure competition}: the presence of the red strain actually changes the sign of the selection coefficient between green and blue (from positive to negative), and the blue strain similarly changes the sign of selection between green and red.  In this example it occurs via modifications to the effective population yield $\Ybar$.  Even if the strains have identical yields, it is possible for the pairwise champion to lose the mixed competition over short time scales due to effects from the growth-lag coupling terms $s_{ijk}^\mathrm{coupling}$ (\emph{Supplementary Methods} Sec.~S7, Fig.~S4).
     

     \textbf{Selection can be non-transitive.}  It is also possible that there is no pairwise champion among a set of strains, meaning that selection is non-transitive~\cite{Verhoef2010}.  For example, in Fig.~\ref{fig:binary_ternary_comparisons}C, red beats blue and green beats red, but then blue beats green, forming a rock-paper-scissors game~\cite{Sinervo1996, Kerr2002}.  This outcome relies crucially on the existence of tradeoffs between growth traits, so that no single growth strategy always wins (\emph{Supplementary Methods} Sec.~S8, Fig.~S5).  In this example, red beats blue by having a shorter lag time, green beats red by growing faster and using resources more efficiently (higher yield), and blue beats green by having shorter lag and hoarding resources (lower yield).  Non-transitivity in this model occurs only for pairwise competitions where each strain starts with equal density ($x_i(0) = 1/2$); invasion competitions, where each strain competes against another starting from very low density (as would occur in an invasion by a migrant or a new mutant), do not demonstrate this type of non-transitivity (\emph{Supplementary Methods} Sec.~S8, Fig.~S5).
     
     
     Non-transitive competitions are particularly confounding for predicting the behavior of a mixed community.  Since there is no clear champion, non-transitive pairwise competitions are often hypothesized as the basis for oscillations or coexistence in mixed communities~\cite{Sinervo1996, Kerr2002, Verhoef2010}.  However, a non-transitive set of strains will not coexist all together in our model.  
Which strain wins, though, is not directly predictable from the pairwise selection coefficients, and in fact may depend on the initial conditions \new{due to frequency-dependent selection}.  For example, Fig.~\ref{fig:binary_ternary_comparisons}D shows the outcomes of ternary competitions for a non-transitive set of strains as a function of their initial densities.  If green starts at sufficiently high density, then it wins the mixed competition, but otherwise red wins.  In the inset we show one such ternary competition, with initial conditions on the boundary between the red and green regimes.  Here the outcome is very sensitive to the initial conditions, since \new{frequency-dependent} higher-order effects from the decaying blue population draw the red and green strains toward their unstable \new{fixed} point, where they temporarily remain until the blue strain goes extinct and either red or green eventually wins.

\section*{Discussion}

     \textbf{Ecological effects of growth trait variation.}  Variation in multiple growth traits is widespread in microbial populations~\cite{Novak2006, Warringer2011, Jasmin2012a}, since even single mutations tend to be pleiotropic with respect to these traits~\cite{Fitzsimmons2010, Adkar2017}.  Genetically-identical cells can also demonstrate significant growth variation~\cite{LevinReisman2010, Ziv2013}.  We have shown how this variation, with competition for only a single finite resource and no other interactions, is sufficient to produce a range of ecological phenomena, such as coexistence, \new{multistability}, keystones, non-transitivity, and other collective behaviors where a community is more than the sum of its parts.  This is because variation in multiple growth traits creates higher-order effects in which the pairwise selection coefficients themselves change in the presence of other strains.  This goes beyond the effects of ordinary clonal interference~\cite{Lang2013}; for example, even the sign of the selection coefficients may change due to these higher-order effects, so that a strain that is the best in pairwise competitions actually goes extinct in the mixed community (Fig.~\ref{fig:binary_ternary_comparisons}A,B).  For example, a mutation that is apparently beneficial against the wild-type alone may not only appear to be less beneficial in the presence of other mutations, but it could even appear to be deleterious.  These results highlight the importance of considering the mutational distribution of ecological effects, rather than just fitness effects relative to a wild-type, for predicting evolutionary dynamics.
     
     The ability to coexist on a single limiting resource contradicts the principle of competitive exclusion~\cite{Hardin1960, Levin1970}.  While previous work indicated that two strains may stably coexist through tradeoffs in growth traits~\cite{Levin1972, Stewart1973, Turner1996, Smith2011, Manhart2018}, here we have shown that an unlimited number of strains can in fact coexist through this mechanism.  \new{Conceptually this is reminiscent of other coexistence mechanisms, such as the storage effect~\cite{Chesson1997}, where tradeoffs in multiple life-history traits allows long-term balancing of competition outcomes.}
Our work also supports the hypothesis that higher-order effects should be widespread in microbial ecosystems~\cite{Billick1994, Momeni2017}.  Experimental tests for these effects and the predictive power of pairwise competitions remains limited, however.  A recent study found that pairwise competitions of soil bacteria generally did predict the behavior of three or more species together~\cite{Friedman2017}, although there were important exceptions.  Our results suggest an avenue for future investigations of this problem.

     \textbf{Tradeoffs between growth traits.}  Tradeoffs among lag, growth, and yield underlie many key outcomes of the model, such as coexistence.  The prevalence of these tradeoffs in microbial populations has been the subject of many previous studies, especially due to interest in the $r$/$K$ (growth-yield) selection problem.  \new{Some models of metabolic constraints do imply a tradeoff between growth rate and yield~\cite{MacLean2007, Pfeiffer2001}, while others propose that both tradeoffs and synergies are possible depending on the environment~\cite{RedingRoman2017}, and experiments have seen evidence of both cases~\cite{Novak2006, Fitzsimmons2010, Warringer2011, Jasmin2012a}.  
     
     The relationship between lag and growth has received less attention.  While models of the lag phase suggest a synergy, rather than a tradeoff, with the growth phase ($c < 0$ in Eq.~\ref{eq:growth_lag_tradeoff})~\cite{Baranyi1998, Baranyi1994, Himeoka2017}, experimental support for this prediction has been mixed.  For example, Ziv et al. found that in a large collection of yeast strains, faster growth mostly corresponded to shorter lag~\cite{Ziv2013, Ziv2017}.  However, other sets of strains in yeast and \emph{E. coli} have found no such trend~\cite{Warringer2011, Adkar2017}.  Quantifying the frequency and strength of these tradeoffs therefore remains an important topic for future investigation.  Regardless of general trends, though, it is clear that growth-lag tradeoffs can be realized within some sets of microbial strains.  For example, the tradeoff was directly observed in \emph{E. coli} strains with certain mutations in adenylate kinase~\cite{Adkar2017}.}     
     
     \textbf{Applications to experimental ecology and evolution.}  
\new{Given a collection of microbial strains and their measured growth traits~\cite{Zwietering1990}, we can in principle use our model to predict the population dynamics of any combination of strains.  If we also know the distribution of mutational effects on growth traits, we can further predict evolutionary dynamics to determine what patterns of traits are likely to evolve, which can be compared with experimental data~\cite{Novak2006, Fitzsimmons2010, Warringer2011, Jasmin2012a}.  In practice, real populations will likely contain more complex interactions beyond competition for a single resource~\cite{Turner1996}, as well as more complex growth dynamics~\cite{Stewart1973}.  Nevertheless, our model provides a valuable tool for interpreting the ecological and evolutionary significance of growth trait variation, especially for generating new hypotheses to be experimentally tested.  For example, it can be used to estimate what role growth trait variation plays in the ecological dynamics of a coexisting community.}

     Our results are especially relevant to laboratory evolution and ecology experiments where populations undergo periodic growth cycles.  While the importance of interference among mutants has been widely studied in these experiments~\cite{Lang2013, Levy2015}, it is generally assumed that each mutant is described by a fixed selection coefficient independent of the background population, since the relative genetic homogeneity of the population suggests there should be no additional ecological interactions beyond competition for the limiting resource.  But since even single mutations will produce variation in multiple growth traits, our results show that higher-order effects should actually be widespread in these populations.  Even genetically-identical populations may experience higher-order effects due to stochastic cell-to-cell variation~\cite{Baranyi1998, LevinReisman2010, Ziv2013}, although the effects will fluctuate from one round of competition to the next assuming cell-to-cell variation does not persist over these timescales.  \new{We look forward to quantifying the importance of these higher-order effects in future work.}

\section*{Methods}


\subsection*{Model of population growth and competition}

     For a population consisting of a single microbial strain, we approximate its growth dynamics by the following minimal model (Fig.~\ref{fig:growth_model}A)~\cite{Buchanan1997}:

\beq
N(t) = \left\{
\begin{array}{ll}
N(0) & 0 \leq t < \lambda, \\
N(0) e^{(t - \lambda)/\tau} & \lambda \leq t < t_\sat, \\
N(0) e^{(t_\sat - \lambda)/\tau} & t \geq t_\sat, \\
\end{array}
\right.
\label{eq:growth_model}
\eeq

\noindent where $\lambda$ is the lag time during which no growth occurs and $\tau$ is the exponential growth time (reciprocal growth rate, or time over which the population grows $e$-fold).  The saturation time $t_\sat$ at which growth stops is determined by the amount of resources in the environment.  We assume that the population size at saturation $N(t_\sat)$ is proportional to the total amount $R$ of the limiting resource.  Let $Y$ denote this constant of proportionality ($N(t_\sat) = RY$), which we will refer to as the intrinsic yield since it is the total number of cells per unit resource~\cite{Vasi1994}.  
Let $\rho = R/N(0)$ be the initial density of resources per cell.  The saturation time then equals

\beq
t_\sat = \lambda + \tau\log\left(\rho Y\right).
\label{eq:tsat_singlestrain}
\eeq

     If there are multiple distinct strains simultaneously competing for the same pool of resources, let each strain $i$ grow according to Eq.~\ref{eq:growth_model} with its own initial frequency $x_i$ and growth traits $\tau_i$, $\lambda_i$, and $Y_i$.  
Since the total amount of resources used by strain $i$ by time $t$ is $N_i(t)/Y_i$, the saturation time $t_\sat$ for the whole population is defined by

\beq
R = \sum_i \frac{N_i(t_\sat)}{Y_i}.
\label{eq:saturation_condition}
\eeq

\noindent By solving this equation for $t_\sat$, we can therefore calculate all properties of the competition, such as the frequencies of each strain at the end.  While we have assumed here that resources are consumed in proportion to the total number of cells, which holds for resources such as space, it is straightforward to modify the model for other modes of resource consumption~\cite{Manhart2018}.  For example, resources may be consumed in proportion to the total number of cell divisions.  The difference in these two models, however, will be negligible if the fold-change of the population over growth is large.


\subsection*{Population dynamics over competition rounds}
     
     If the population undergoes many rounds of dilution and resource renewal (Fig.~\ref{fig:growth_model}B), the density of strain $i$ at the end of a round equals its density at the beginning of the next round (ignoring stochastic effects of sampling).  Let $x_i(r)$ be the density of strain $i$ at the beginning of competition round $r$ and $x_i'(r)$ be the density at the end, so that $x_i'(r) = x_i(r+1)$.  The selection coefficients determine how the densities change over the round.   Using the selection coefficient definition $s_{ij} = \log(x_i'(r)/x_j'(r)) - \log(x_i(r)/x_j(r))$ (Eq.~\ref{eq:s_def}), we can obtain the recurrence relation for the change in density over each round:

\beq
x_i(r+1) = \frac{x_i(r)}{\sum_k x_k(r)e^{s_{ki}(\mathbf{x}(r))}},
\label{eq:freq_dynamics}
\eeq

\noindent where $\mathbf{x}(r)$ is the vector of densities $\{x_k(r)\}$ at the beginning of round $r$.  Over a large number of rounds, these dynamics can be approximated by a differential equation:

\beq
\begin{split}
\frac{dx_i}{dr} & = \frac{x_i}{\sum_k x_k e^{s_{ki}(\mathbf{x})}} - x_i \\
& \approx x_i \sum_k x_k s_{ik}(\mathbf{x}), \\
\end{split}
\label{eq:freq_ODE}
\eeq

\noindent where on the second line we have invoked the approximation that all $s_{ki}$ values are small.  This is of Lotka-Volterra form where the selection coefficients encode the effective (density-dependent) interaction coefficients between strains.


\begin{acknowledgments}
     We thank Tal Einav for his detailed comments on the manuscript.  This work was supported by NIH awards F32 GM116217 to MM and R01 GM068670 to EIS.
\end{acknowledgments}


\section*{Author contributions}

     MM and EIS designed research; MM performed calculations; MM wrote the manuscript.  Both authors edited and approved the final version.


\bibliographystyle{unsrt}
\bibliography{References}


\end{document}


\title{Supplementary Methods: \\ Growth tradeoffs produce complex microbial communities \new{on a single limiting resource}}

\author{Michael Manhart}
\affiliation{Department of Chemistry and Chemical Biology, Harvard University, Cambridge, MA 02138, USA}
\author{Eugene I. Shakhnovich}
\affiliation{Department of Chemistry and Chemical Biology, Harvard University, Cambridge, MA 02138, USA}
\date{\today}

\maketitle

\setcounter{page}{1}
\setcounter{equation}{0}
\setcounter{figure}{0}
\setcounter{section}{0}
\setcounter{subsection}{0}
\renewcommand{\theequation}{S\arabic{equation}}
\renewcommand{\thefigure}{S\arabic{figure}}
\renewcommand{\thesection}{S\arabic{section}}


\section{Derivation of the selection coefficient}

     To calculate an explicit formula for the selection coefficients as functions of the underlying parameters, we assume the nontrivial case in which the saturation time is longer than each strain's lag time ($t_\sat > \max_k \lambda_k$).  Then using the minimal growth model in Eq.~\eqgrowthmodel{}, the selection coefficient definition in Eq.~\eqsdef{} simplifies to
     
\beq
s_{ij} = \frac{1}{\tau_i}(t_\sat - \lambda_i) - \frac{1}{\tau_j}(t_\sat - \lambda_j).
\eeq

\noindent We next rewrite the saturation condition (Eq.~\eqsaturationcondition{}) in terms of the selection coefficients relative to strain $i$:
     
\beq
1 = e^{(t_\sat - \lambda_i)/\tau_i} \left(\sum_k \frac{x_k}{\rho Y_k} e^{s_{ki}} \right).
\eeq

\noindent We can then solve for $t_\sat$ and expand to first order in each $s_{ki}$:

\beq
\begin{split}
t_\sat = & ~\lambda_i - \tau_i \log\left(\sum_k \frac{x_k}{\rho Y_k} e^{s_{ki}} \right) \\
\approx & ~\lambda_i + \tau_i \left(\log\left[\rho\Ybar\right] - \Ybar \sum_k \frac{x_k}{Y_k} s_{ki} \right), \\
\end{split}
\label{eq:tsat_vs_s}
\eeq

\noindent where $\Ybar = \left(\sum_k x_k/Y_k\right)^{-1}$ is the harmonic mean of the yields (Eq.~\eqtYbar{}).  However, since we can freely choose a different strain $j$ to be the reference strain, we must also have

\beq
t_\sat \approx \lambda_j + \tau_j \left(\log\left[\rho\Ybar\right] - \Ybar \sum_k \frac{x_k}{Y_k} s_{kj} \right).
\eeq

\noindent To be self-consistent these two expressions for $t_\sat$ must be equal for any $i$ and $j$, which leads to the following system of linear equations for the selection coefficients:

\begin{multline}
s_{ij} - \Ybar \frac{\Delta\tau_{ij}}{\tau_i} \sum_k \frac{x_k}{Y_k} s_{kj} = \\
- \frac{1}{\tau_{i}} \left(\Delta\tau_{ij} \log\left[\rho\Ybar\right] + \Delta\lambda_{ij}\right),
\label{eq:s_system}
\end{multline}

\noindent where $\Delta\lambda_{ij} = \lambda_i - \lambda_j$ and $\Delta\tau_{ij} = \tau_i - \tau_j$.




     We now take the solution for $s_{ij}$ in Eq.~\eqsformula{} as an ansatz and show that it satisfies this system of equations.  If we substitute Eq.~\eqsformula{} for $s_{ij}$ and $s_{kj}$ in Eq.~\ref{eq:s_system}, the left-hand side (LHS) of the system in Eq.~\ref{eq:s_system} becomes
     
\begin{widetext}
\begin{multline}
\mathrm{LHS} = -\frac{\tbar}{\tau_i\tau_j}\left(\Delta\tau_{ij} \log\left[\rho\Ybar\right] + \Delta\lambda_{ij} + \Ybar\sum_k \frac{x_k}{Y_k\tau_k}[\Delta\tau_{ik}\Delta\lambda_{kj} - \Delta\lambda_{ik}\Delta\tau_{kj}] \right) \\ 
+ \Ybar\frac{\Delta\tau_{ij}}{\tau_i} \sum_k \frac{x_k}{Y_k} \frac{\tbar}{\tau_j\tau_k} \left(\Delta\tau_{kj} \log\left[\rho\Ybar\right] + \Delta\lambda_{kj} + \Ybar \sum_\ell \frac{x_\ell}{Y_\ell\tau_\ell}[\Delta\tau_{k\ell}\Delta\lambda_{\ell j} - \Delta\lambda_{k\ell}\Delta\tau_{\ell j}] \right).
\end{multline}
\end{widetext}
     
\noindent Since 

\beq
\sum_k \sum_\ell \frac{x_k}{Y_k\tau_k} \frac{x_\ell}{Y_\ell\tau_\ell} [\Delta\tau_{k\ell}\Delta\lambda_{\ell j} - \Delta\lambda_{k\ell}\Delta\tau_{\ell j}] = 0
\label{eq:antisymmetric_summand}
\eeq

\noindent because the summand is antisymmetric in the summation indices, we drop the inner sum over $\ell$ and combine the remaining sums over $k$ to obtain


\begin{multline}
\mathrm{LHS} = -\frac{\tbar}{\tau_i\tau_j}\left(\Delta\tau_{ij} \log\left[\rho\Ybar\right] + \Delta\lambda_{ij} + \Ybar \sum_k \frac{x_k}{Y_k\tau_k} \right. \\
\times \left[\Delta\tau_{ik}\Delta\lambda_{kj} - \Delta\lambda_{ik}\Delta\tau_{kj} \right. \\
\left. \phantom{\sum_k} \left. - \Delta\tau_{ij}\Delta\tau_{kj} \log\left(\rho\Ybar\right) - \Delta\tau_{ij}\Delta\lambda_{kj}\right] \right).
\end{multline}

\noindent We cancel out terms and factor to obtain

\beq
\begin{split}
\mathrm{LHS} & = -\frac{\tbar}{\tau_i\tau_j} (\Delta\tau_{ij} \log\left[\rho\Ybar\right] + \Delta\lambda_{ij}) \left(1 - \Ybar\sum_k \frac{x_k}{Y_k\tau_k} \Delta\tau_{kj} \right) \\
& = -\frac{\tbar}{\tau_i\tau_j} (\Delta\tau_{ij} \log\left[\rho\Ybar\right] + \Delta\lambda_{ij}) \left(1 - \left[1 - \frac{\tau_j}{\tbar}\right] \right) \\
& = - \frac{1}{\tau_i} (\Delta\tau_{ij} \log\left[\rho\Ybar\right] + \Delta\lambda_{ij}), \\
\end{split}
\eeq

\noindent where we have used the definitions in Eq.~\eqtYbar{} to invoke 

\beq
\Ybar \sum_k \frac{x_k}{\tau_k Y_k} \Delta\tau_{kj} = 1 - \frac{\tau_j}{\tbar}.
\label{eq:HT_identity}
\eeq

\noindent This equals the right side of Eq.~\ref{eq:s_system}, proving the solution is correct.


\section{Saturation time and overall yield for a mixed population}

     Here we calculate expressions for the saturation time and overall yield for a mixed population of multiple strains, which provide interpretations of the quantities $\tbar$ and $\Ybar$ in Eq.~\eqtYbar{}.  Using the approximation for $t_\sat$ in Eq.~\ref{eq:tsat_vs_s}, which holds for any reference strain $i$, we insert the selection coefficients $s_{ki}$ from the explicit formula in Eq.~\eqsformula{}:

\begin{widetext} 
\beq
\begin{split}
t_\sat & \approx \lambda_i + \tau_i\left(\log\left[\rho\Ybar\right] - \Ybar\sum_k \frac{x_k}{Y_k} s_{ki}\right) \\
& = \lambda_i + \tau_i\log\left(\rho\Ybar\right) + \tau_i \Ybar\sum_k \frac{x_k}{Y_k} \frac{\tbar}{\tau_k\tau_i} \left(\Delta\tau_{ki} \log\left[\rho\Ybar\right] + \Delta\lambda_{ki} + \Ybar \sum_\ell \frac{x_\ell}{Y_\ell \tau_\ell} [\Delta\tau_{k\ell} \Delta\lambda_{\ell i} - \Delta\lambda_{k\ell} \Delta\tau_{\ell i}] \right). \\
\end{split}
\eeq
\end{widetext}

\noindent We eliminate the double sums over $k$ and $\ell$ (since the summand is antisymmetric in the summation indices, as in Eq.~\ref{eq:antisymmetric_summand}) and invoke the identity in Eq.~\ref{eq:HT_identity} to obtain

\beq
\new{t_\sat \approx \sum_k x_k \lambda_k \frac{\tbar\Ybar}{\tau_kY_k}  + \tbar\log\left(\rho\Ybar\right).}
\label{eq:tsat_approx}
\eeq

\noindent Comparing with the saturation time for a homogeneous population of a single strain (Eq.~\eqtsatsinglestrain{}), we see the weighted sum over all lag times corresponds to the effective time shift from the lag phase, while the last term determines the time during which exponential growth occurs.  In analogy with Eq.~\eqtsatsinglestrain{}, $\tbar$ is therefore the mixed population's effective exponential growth time (reciprocal growth rate), and $\Ybar$ is the mixed population's effective yield.  Note that the effective growth rate $1/\tbar$ is just an arithmetic mean of the individual strains' growth rates.

     The quantity $\Ybar$ is in fact the exact yield for a mixed population when all strains are neutral (coexisting).  Let $N_\sat = \sum_k N_k(t_\sat)$ be the total population size at saturation. 
For a set of strains to all be neutral, they must have a fixed growth-lag tradeoff $c_{ij} = -\Delta\lambda_{ij}/\Delta\tau_{ij} = c$ for all pairs of strains $i$ and $j$ (Eq.~\eqgrowthlagtradeoff{}).  In that case the effective yield is $\Ybar = e^c/\rho$ (Sec.~\ref{sec:coexistence}).  Since all $s_{ij} = 0$ by definition, then Eq.~\ref{eq:tsat_vs_s} implies that $t_\sat = \lambda_k + c\tau_k$ for any strain $k$, which we can rewrite as $(t_\sat - \lambda_k)/\tau_k = c$.  We then calculate the total population size at saturation to be
     
\beq
\begin{split}
N_\sat & = \sum_k N_k(0) e^{(t_\sat - \lambda_k)/\tau_k} \\
& = e^c \sum_k N_k(0) \\
& = \Ybar \rho \sum_k N_k(0). \\
\end{split}
\eeq

\noindent Since $\rho \sum_k N_k(0)$ equals the total amount of resources $R$, this shows that the total saturation size $N_\sat$ is proportional to $R$ with constant $\Ybar$, meaning that $\Ybar$ is indeed the yield for the whole population.

\section{Conditions for coexistence}
\label{sec:coexistence}

     Here we derive the conditions on parameters for multiple strains to coexist, i.e., all pairwise selection coefficients are zero.  We therefore substitute $s_{ij} = 0$ into the linear system for the selection coefficients (Eq.~\ref{eq:s_system}).  This implies the right-hand side of the system must be zero: $\Delta\tau_{ij} \log\left(\rho\Ybar\right) + \Delta\lambda_{ij} = 0$, or equivalently $\rho\Ybar = e^{-\Delta\lambda_{ij}/\Delta\tau_{ij}}$.  Since this must hold for all pairs of strains $i$ and $j$, $\Delta\lambda_{ij}/\Delta\tau_{ij}$ is therefore a constant independent of $i$ and $j$.  We thus arrive at the condition for coexistence:
     
\beq
\rho\Ybar = e^{c}
\label{eq:coexistence}
\eeq

\noindent where 

\beq
c = -\frac{\Delta\lambda_{ij}}{\Delta\tau_{ij}}
\label{eq:c_def}
\eeq

\noindent for all pairs of strains $i$ and $j$, which is equivalent to the linear growth-lag tradeoff condition in Eq.~\eqgrowthlagtradeoff{}.  Equation~\ref{eq:coexistence} implies that $c > 0$ (growth and lag must have a tradeoff, rather than a synergy, across strains), since the left-hand side of Eq.~\ref{eq:coexistence} is the fold-change of the whole population's growth and therefore must be greater than 1.  Equation~\ref{eq:coexistence} furthermore imposes a constraint on the initial density of resources $\rho$.  Since $\Ybar$ is the harmonic mean of the yields $\{Y_k\}$ (Eq.~\eqtYbar{}), it is bounded by the minimum and maximum yields:
     
\beq
\min_k Y_k < \Ybar < \max_k Y_k.
\eeq

\noindent Combining this constraint with Eq.~\ref{eq:coexistence}, we obtain the limits on the resource density $\rho$ in Eq.~\eqrhobounds{}.


\subsection{Space of densities with coexistence}

\begin{figure}
\centering\includegraphics[width=\columnwidth]{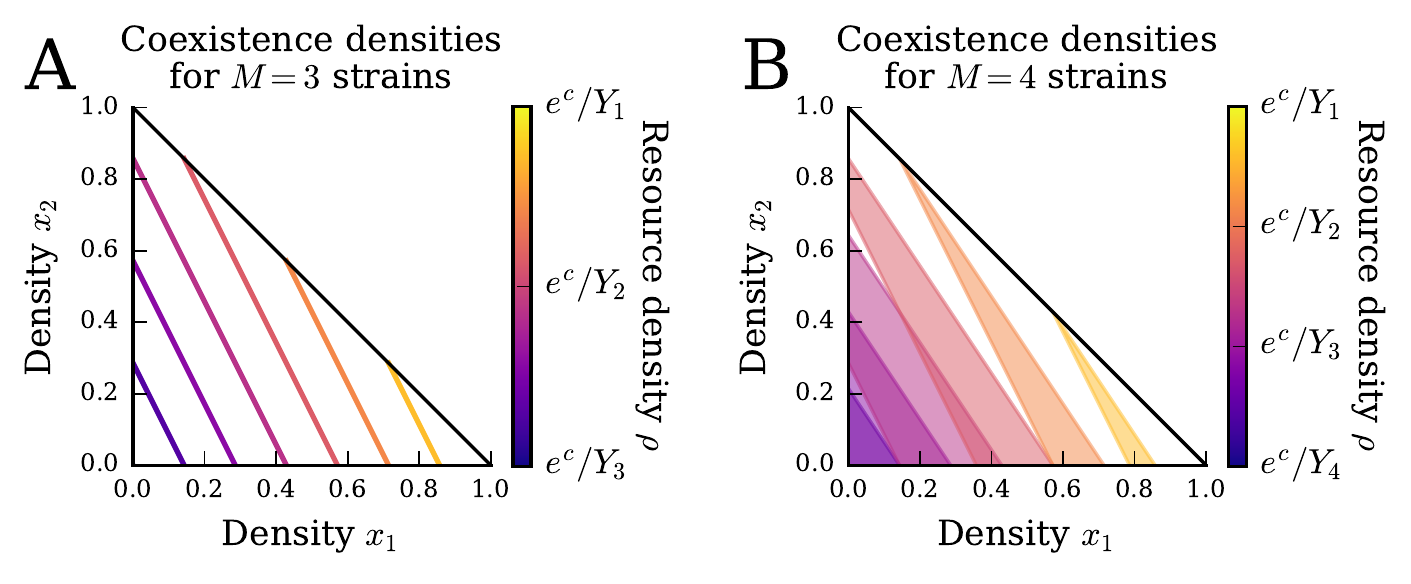}
\caption{
\textbf{Space of densities at which strains coexist.}  
(A)~For three strains, the space of coexistence densities (satisfying Eq.~\ref{eq:coexistence_freqs}) is one-dimensional.  We project this one-dimensional region into the space of densities for two strains $x_1$ and $x_2$, with different possible values of the resource density $\rho$ (indicated by the color).  
(B)~Same as (A) but for four strains, where the space of coexistence densities is two-dimensional.  
See Sec.~\ref{sec:parameters} for parameter values.
}
\label{fig:coexistence_freq_space}
\end{figure}

     The condition Eq.~\ref{eq:coexistence} also imposes constraints on the densities at which coexistence can occur, provided the strains' traits satisfy the growth-lag tradoeff (Eq.~\eqgrowthlagtradeoff{}) and the resource density is within the necessary bounds (Eq.~\eqrhobounds{}).  Substituting in the definition of $\Ybar$ to Eq.~\ref{eq:coexistence}, a set of densities $\{\tilde{x}_k\}$ at which the strains coexist must satisfy the following linear equation:

\beq
\sum_k \frac{\tilde{x}_k}{Y_k} = \rho e^{-c}.
\label{eq:coexistence_freqs}
\eeq

\noindent If there are $M$ total strains, then the space of coexistence densities is a section of an $(M-2)$-dimensional hyperplane, since the $M$ densities must satisfy two linear equations (Eq.~\ref{eq:coexistence_freqs} as well as normalization $\sum_k \tilde{x} = 1$).  Figure~\ref{fig:coexistence_freq_space} shows these strain densities for $M = 3$ (Fig.~\ref{fig:coexistence_freq_space}A) and $M = 4$ strains (Fig.~\ref{fig:coexistence_freq_space}B) as functions of the resource density $\rho$, which determines which strains are more highly represented in the coexisting community.


\subsection{Coexistence with maximum entropy over strains}

     The set of coexistence densities $\{\tilde{x}_k\}$ with maximum diversity over strains is of particular interest.  A common way to measure diversity in a population is by Shannon entropy, defined as 
     
\beq
S = - \sum_k \tilde{x}_k \log \tilde{x}_k.
\eeq

\noindent This ranges from zero if only one strain is present, to $\log M$ if all $M$ strains are equally abundant.  The coexistence condition on densities (Eq.~\ref{eq:coexistence_freqs}) means that the reciprocal yield averaged over densities must be $\rho e^{-c}$.  With this constraint the maximum-entropy set of densities is of Boltzmann form~\cite{Presse2013}, with reciprocal yield in the role of energy:

\beq
\tilde{x}_k = \frac{1}{Z} e^{-\beta/Y_k},
\label{eq:max_ent_freqs}
\eeq

\noindent where $\beta$ is defined such that

\beq
\frac{\sum_k \frac{1}{Y_k} e^{-\beta/Y_k}}{\sum_k e^{-\beta/Y_k}} = \rho e^{-c}
\eeq

\noindent and $Z = \sum_k e^{-\beta/Y_k}$ is the normalization constant.  The maximum entropy is therefore

\beq
S_{\max} = \beta\rho e^{-c} + \log\left(\sum_k e^{-\beta/Y_k}\right).
\eeq

     The parameter $\beta$ (analogous to inverse temperature) sets a yield threshold determining how much of the population consists of strains with low yields versus those with high yields.  If $\rho$ is close to $e^c/\min_k Y_k$ (high end of range in Eq.~\eqrhobounds{}), then $\beta$ will be negative and large in magnitude, meaning the strains with the lowest yields will be favored.  Similarly, if $\rho$ is close to $e^c/\max_k Y_k$ (low end of range in Eq.~\eqrhobounds{}), then $\beta$ will be large and positive, so that strains with the highest yields will be favored.  All strains will be equally represented ($x_k = 1/M$) if $\beta = 0$, which occurs if the resource density is set to
     
\beq
\rho = \frac{e^c}{M}\sum_k \frac{1}{Y_k}.
\eeq


\subsection{Effective coexistence with noisy tradeoffs}

\begin{figure}
\centering\includegraphics[width=0.8\columnwidth]{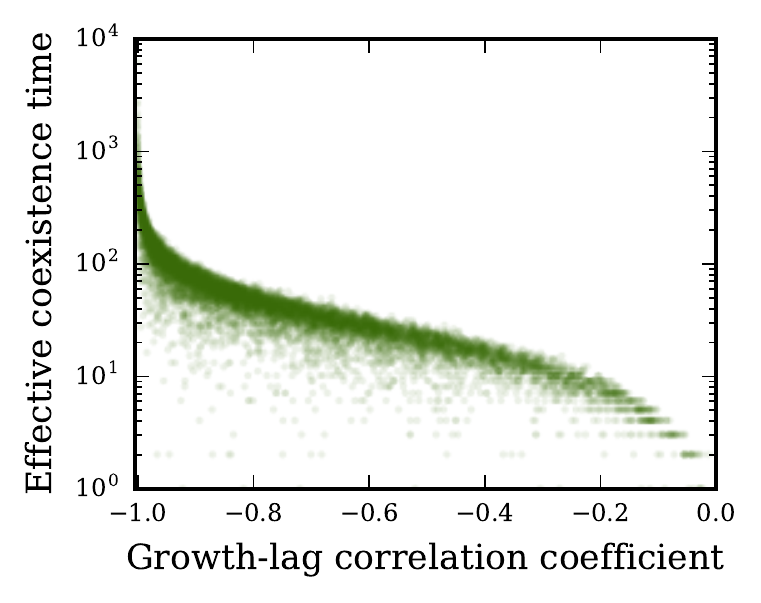}
\caption{
\textbf{Effective coexistence of strains with approximate growth-lag tradeoffs.}   
Each green point represents a set of 100 strains with randomly-generated trait values.  We sample the yields $\{Y_k\}$ from a Gaussian distribution with mean $10^3$ and standard deviation $10^2$.  We sample growth times $\{\tau_k\}$ from a Gaussian with mean $1$ and standard deviation $10^{-2}$; we then generate correlated lag times $\{\lambda_k\}$ from the growth times using a ``true'' correlation coefficient uniformly sampled between $-1$ and $0$.  The resource density is $\rho = 1$.  From the resulting set of 100 strains, we determine the empirical growth-lag tradeoff $c$ and correlation coefficient from the linear regression of growth and lag times.  The population starts at the maximum-entropy set of densities that would allow coexistence (Eq.~\ref{eq:max_ent_freqs}) if the strains' growth and lag times fell exactly on the regression line.  We exclude realizations where $\rho = 1$ falls outside of the allowed range (Eq.~\eqrhobounds{}) given the tradeoff $c$ and yields $\{Y_k\}$, or if any of the maximum-entropy densities is too small ($< 10^{-6}$) to constitute meaningful coexistence.  We then evolve the densities over competition rounds (Eq.~\eqfreqdynamics{}) and determine the number of rounds until any strain goes extinct (i.e., its density falls below $10^{-6}$), which we use as the time of effective coexistence.  \new{For each of the 100 communities, this time is plotted on the vertical axis against the empirical growth-lag correlation coefficient on the horizontal axis.}  
Green points are transparent to show their density.
}
\label{fig:coexistence_robustness}
\end{figure}

     The condition for coexistence in Eq.~\eqgrowthlagtradeoff{} is an exact linear tradeoff between growth and lag times.  A tradeoff among real strains will never be exactly linear for more than two strains, however.  If the tradeoff is noisy, with some fluctuations around a linear trend, then this will still lead to effective coexistence over some finite time scale, which may be sufficiently long to be biologically relevant \new{(e.g., to observe in a laboratory experiment, or before new mutations arise or the environment changes)}.  To illustrate this, we randomly generate communities (sets of strains with distinct growth traits) with different correlations of growth and lag times across strains.  We initialize each community at a set of densities $\{x_k\}$ such that it would coexist if all the traits exactly obeyed the linear regression of growth traits.  We then measure the time (number of competition rounds) it takes for the first strain to go extinct, i.e., for its density to drop below a certain threshold.  Figure~\ref{fig:coexistence_robustness} shows that while strains with only weak growth-lag correlations will generally not coexist for very long, as expected, the apparent coexistence time increases rapidly as the correlation becomes stronger.  Thus, a community with even moderate correlation may still practically coexist over a significant time.


\section{Stability of coexistence to density fluctuations}

\begin{figure*}
\centering\includegraphics[width=\textwidth]{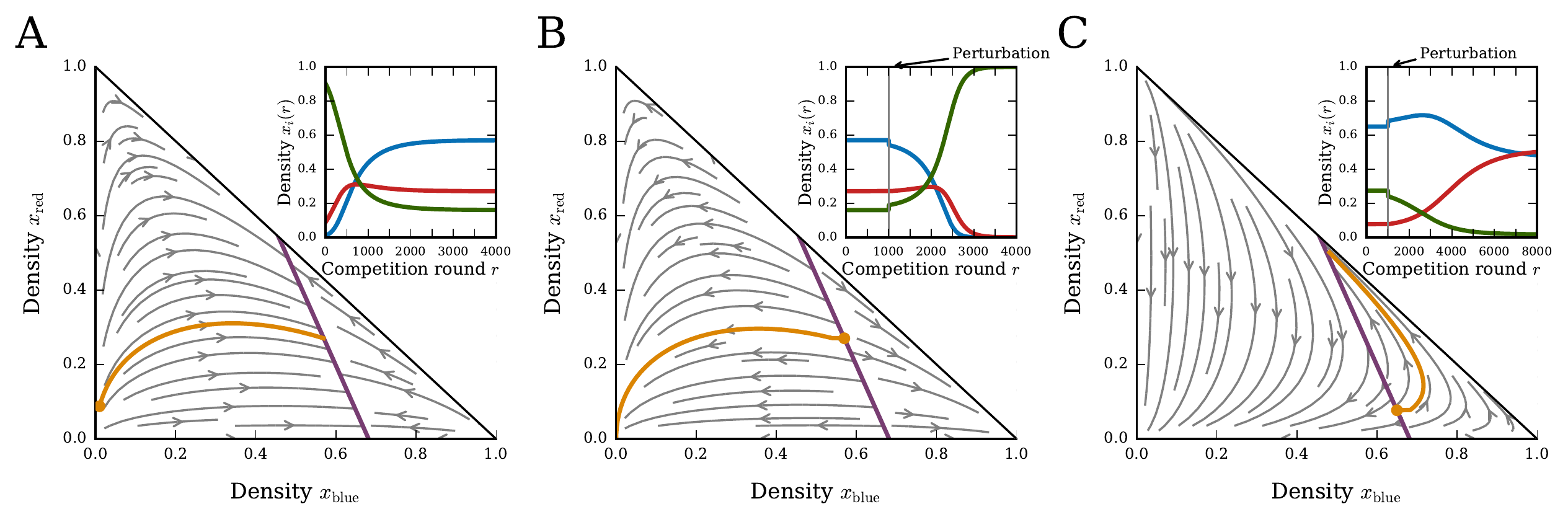}
\caption{
\textbf{Stability of coexistence.}  
Dynamics of three strains (blue, red, green) projected into the space of densities $x_\blue$ and $x_\red$.  Gray streamlines show $dx_i/dr$ (Eq.~\eqfreqODE{}), while the magenta line indicates the set of coexistence densities (Eq.~\ref{eq:coexistence_freqs}).  The orange curve is an example trajectory beginning at the orange dot.  Insets show the density $x_i(r)$ over competition rounds $r$ for each strain along the orange trajectory.    
(A)~Case where all coexistence densities are stable \new{to perturbations off the coexistence region}; 
(B)~case where all coexistence densities are unstable; 
(C)~case where there is a mix of stable and unstable sets of densities.  
The vertical gray lines in the insets of panels (B) and (C) indicate the time at which the densities are perturbed away from unstable coexistence.  See Sec.~\ref{sec:parameters} for parameter values.
}
\label{fig:stability}
\end{figure*}

     Density fluctuations will occur from both extrinsic and intrinsic noise, such as the random sampling of the population from one round of competition to the next.  Let $\tilde{\mathbf{x}} = \{\tilde{x}_k\}$ be a set of densities satisfying coexistence (Eq.~\ref{eq:coexistence_freqs}).  To determine the stability, we consider small perturbations $\Delta\mathbf{x}$ around this point.  Let the Jacobian of the selection coefficients at the coexistence point be

\beq
\mathcal{J}_{ijk} = \left. \frac{\partial s_{ij}}{\partial x_k} \right|_{\mathbf{x} = \tilde{\mathbf{x}}}
\eeq

\noindent so that

\beq
s_{ij}(\tilde{\mathbf{x}} + \Delta\mathbf{x}) \approx \sum_k \mathcal{J}_{ijk} \Delta x_k.
\eeq

\noindent For small perturbations we can approximate the density dynamics with the differential equations in Eq.~\eqfreqODE{}.  Therefore the dynamics around the coexistence point at competition round $r$ are

\beq
\begin{split}
\frac{d}{dr} \Delta x_i & = -(\tilde{x}_i + \Delta x_i) \sum_k (\tilde{x}_k + \Delta x_k) s_{ki}(\tilde{\mathbf{x}} + \Delta\mathbf{x}) \\
& \approx -\tilde{x}_i \sum_k \tilde{x}_k \sum_j \mathcal{J}_{kij} \Delta x_j, \\
\end{split}
\eeq

\noindent where we have dropped higher-order terms in $\Delta\mathbf{x}$.

     To analyze the stability, we must obtain the eigenvalues of the matrix

\beq
W_{ij} = -\tilde{x}_i \sum_k \tilde{x}_k \mathcal{J}_{kij},
\eeq

\noindent since negative eigenvalues will correspond to directions in the space of densities that are stable to small perturbations, positive eigenvalues will indicate unstable directions, \new{and zero eigenvalues indicate neutral directions}.  We calculate the Jacobian of the selection coefficient using the formula in Eq.~\eqsformula{}.  We first note that the density derivatives of $\tbar$ and $\Ybar$ are

\beq
\frac{\partial\tbar}{\partial x_k} = \tbar\frac{\Ybar}{Y_k}\left(1 - \frac{\tbar}{\tau_k}\right), \qquad \frac{\partial\Ybar}{\partial x_k} = -\frac{\Ybar^2}{Y_k}.
\eeq

\noindent Therefore the derivatives of the selection coefficient are

\beq
\begin{split}
\frac{\partial}{\partial x_k} s_{ij}^\mathrm{lag} & = -\frac{\tbar\Delta\lambda_{ij}}{\tau_i\tau_j} \frac{\Ybar}{Y_k} \left(1 - \frac{\tbar}{\tau_k}\right), \\
\frac{\partial}{\partial x_k} s_{ij}^\mathrm{growth} & = -\frac{\tbar\Delta\tau_{ij}}{\tau_i\tau_j} \frac{\Ybar}{Y_k} \left( \left[1 - \frac{\tbar}{\tau_k}\right]\log\left[\rho\Ybar\right] - 1 \right), \\
\frac{\partial}{\partial x_k} s_{ij\ell}^\mathrm{coupling} & = - \frac{\tbar\Ybar}{\tau_\ell Y_\ell} \left(\frac{\Delta\tau_{i\ell}\Delta\lambda_{\ell j} - \Delta\lambda_{i\ell}\Delta\tau_{\ell j}}{\tau_i \tau_j}\right) \\
& \qquad \times \left(\delta_{\ell k} - x_\ell \frac{\tbar\Ybar}{\tau_k Y_k} \right).
\end{split}
\eeq

\noindent Summing these components and evaluating at densities $\{\tilde{x}_k\}$ satisfying coexistence (Eqs.~\ref{eq:coexistence} and~\ref{eq:c_def}) results in

\beq
\mathcal{J}_{ijk} = \frac{\tbar\Delta\tau_{ij}}{\tau_i\tau_j} \frac{\Ybar}{Y_k}.
\eeq

\noindent Therefore the matrix for dynamics around coexistence is 

\beq
\begin{split}
W_{ij} & = -\tilde{x}_i \sum_k \tilde{x}_k \frac{\tbar\Delta\tau_{ki}}{\tau_k\tau_i} \frac{\Ybar}{Y_j} \\
& = -\tilde{x}_i\frac{\tbar\Ybar}{\tau_i Y_j} \left(1 - \tau_i \sum_k \frac{\tilde{x}_k}{\tau_k}\right). \\
\end{split}
\eeq

\noindent This matrix has outer-product form $W_{ij} = a_i b_j$.  It is straightforward to show that such a matrix has one eigenvalue $\mu = \sum_i a_i b_i$, while all other eigenvalues are zero.  The zero eigenvalues correspond to \new{neutral} directions within the region of coexistence, while the one nonzero eigenvalue is the only direction orthogonal to the region.  This eigenvalue is

\beq
\begin{split}
\mu & = - \sum_i \tilde{x}_i\frac{\tbar\Ybar}{\tau_i Y_i} \left(1 - \tau_i \sum_k \frac{\tilde{x}_k}{\tau_k}\right) \\
& = \tbar\sum_k \frac{\tilde{x}_k}{\tau_k} - 1, \\
\end{split}
\eeq

\noindent where we have simplified using the definitions of $\tbar$ and $\Ybar$ (Eq.~\eqtYbar{}).

     Coexistence is stable \new{in the non-neutral direction} when $\mu < 0$, which we can rewrite as

\beq
\sum_k \frac{\tilde{x}_k}{Y_k \tau_k} - \left(\sum_k \frac{\tilde{x}_k}{Y_k}\right) \left(\sum_k \frac{\tilde{x}_k}{\tau_k} \right) > 0.
\eeq

\noindent That is, coexistence is stable when the covariance of reciprocal growth times and reciprocal yields is positive.  If slower-growing strains always have higher yields (growth-yield tradeoff), the correlation is positive for any set of densities $\{\tilde{x}_k\}$, and so any perturbations off the coexistence region will be stabilized.  Figure~\ref{fig:stability}A shows an example with such a tradeoff for three strains; any initial state of the population inevitably converges to coexistence.  If slower-growing strains always have lower yields, then the correlation is negative for any densities, and coexistence will be unstable (Fig.~\ref{fig:stability}B).  If no perfect correlation holds across the growth times and yields, then the sign of the covariance may depend on the densities $\{\tilde{x}_k\}$, leading to a mix of stable and unstable points across the coexistence region as in Fig.~\ref{fig:stability}C.  In that case it is possible for a population that is perturbed away from an unstable set of coexistence densities to evolve to a different, stable set of coexistence densities.


\section{Selection coefficient of an invader relative to a coexisting community}

     Consider a strain that invades a community of coexisting strains.  We assume the invader enters at infinitesimally-low density.  In this case, the quantities $\tbar$ and $\Ybar$ (Eq.~\eqtYbar{}) for the resident strains combined with the invader are essentially the same as their values for the resident strains alone.  In particular, $\log\left(\rho\Ybar\right) \approx c$ (Eq.~\ref{eq:coexistence}), where $c$ is the growth-lag tradeoff for the resident strains.  The selection coefficient of the invader relative to each resident strain $j$ is therefore

\begin{widetext}
\beq
\begin{split}
s_{\mathrm{inv}, j} & = -\frac{\tbar}{\tau_\mathrm{inv}\tau_j}\left(\Delta\lambda_{\mathrm{inv},j} + \Delta\tau_{\mathrm{inv},j} \log\left[\rho\Ybar\right] + \Ybar \sum_k \frac{x_k}{\tau_k Y_k} [\Delta\tau_{\mathrm{inv},k} \Delta\lambda_{kj} - \Delta\lambda_{\mathrm{inv},k} \Delta\tau_{kj}] \right) \\
& \approx -\frac{\tbar}{\tau_\mathrm{inv}\tau_j} \left(\Delta\lambda_{\mathrm{inv}, j} + c\Delta\tau_{\mathrm{inv},j}\right) \left(1 - \frac{e^c}{\rho} \sum_{k \neq \mathrm{inv}} \frac{x_k}{\tau_k Y_k} \Delta\tau_{kj}\right) \\
& = -\frac{1}{\tau_\mathrm{inv}} \left(\Delta\lambda_{\mathrm{inv}, j} + c\Delta\tau_{\mathrm{inv},j}\right) \\
& = -\frac{1}{\tau_\mathrm{inv}} \left(c\tau_\mathrm{inv} + \lambda_\mathrm{inv} - \mathrm{constant}\right), \\
\end{split}
\eeq
\end{widetext}

\noindent where we have used $\lambda_j = -c\tau_j + \mathrm{constant}$ (Eq.~\eqgrowthlagtradeoff{}) for the coexisting strains.  This shows that the selection coefficients between the invader and each of the resident strains $j$ are the same: if the invader's growth and lag times are above the diagonal line of the resident strains (Fig.~\figcoexistence{}A), then the the invader will have negative selection coefficient relative to all resident strains, and so its densities will decay to zero.  Otherwise, its selection coefficient will be positive and the invader will take over.
     

\section{Coexistence of pairs within a community}

     A collection of strains that coexist in pairs will only coexist all together if they share the same growth lag tradeoff $c$ (Eq.~\eqgrowthlagtradeoff{}).  In this case, though, the resource densities $\rho$ at which each pair coexists will not be the same, nor will they be the same as the values of $\rho$ at which the whole community will coexist.  Consider three strains with a growth lag tradeoff $c$ and in order of increasing yields, so that $e^c/Y_3 < e^c/Y_2 < e^c/Y_1$.  For strains 1 and 2 to coexist, $\rho$ must be between $e^c/Y_2$ and $e^c/Y_1$, while for strains 2 and 3 to coexist, $\rho$ must be between $e^c/Y_3$ and $e^c/Y_2$.  These constraints are mutually exclusive, so strains 1 and 2 will not coexist in the same environmental conditions as strains 2 and 3.  Furthermore, all three strains can coexist as long as $e^c/Y_3 < \rho < e^c/Y_1$, but for any value of $\rho$ in that range, one of the pairs will not coexist.  Therefore some, but not all, pairs of strains from a coexisting community will coexist on their own in the same environment.
     

\section{Pairwise champion must win mixed competition with equal yields}

     Figure~\figbinaryternarycomparisons{}A,B gives an example of strains where the pairwise champion (green strain, which wins each binary competition) does not necessarily win the ternary competition with all strains present.  However, that outcome requires the three strains to have significantly different yields.  Here we show that if the strains have equal yields, then the pairwise champion must always win the mixed competition, although it can still lose on short time scales.

     Define the signed component of the selection coefficient to be 
     
\beq
\sigma_{ij} = \frac{\tau_i\tau_j}{\tbar} s_{ij}.
\eeq

\noindent That is, we remove the overall factor of $\tbar/(\tau_i\tau_j)$ from $s_{ij}$ (Eq.~\eqsformula{}) since it is always positive and therefore does not affect the overall sign.  For a set of strains with equal yields, these signed components are convenient because their values for the mixed competition $\sigma_{ij}^\mathrm{mixed}$, where all strains are present, have a simple relationship to their values for binary competitions, $\sigma_{ij}^\mathrm{binary}$, where only $i$ and $j$ are present:

\begin{multline}
\sigma_{ij}^\mathrm{mixed} = (x_i + x_j)\sigma_{ij}^\mathrm{binary} \\
+ \sum_{k \neq i,j} \frac{x_k}{\tau_k} \left(\tau_j\sigma_{ik}^\mathrm{binary} - \tau_i\sigma_{jk}^\mathrm{binary}\right).
\label{eq:binary_to_mixed}
\end{multline}

\noindent That is, the selection coefficient on $i$ relative to $j$ in the mixed competition is a linear combination of the selection coefficient from their binary competition, weighed by the fraction of the mixed population consisting of $i$ and $j$, with the binary selection coefficients of $i$ and $j$ relative to all other strains $k$, each weighed by the density of that other strain.
     
     In the case of equal yields, selection coefficients for binary competitions must obey transitivity~\cite{Manhart2018}, and therefore there must be one strain that wins all of the binary competitions, and another strain that loses all of them.  If $i$ is the winner of all binary competitions ($\sigma_{ik}^\mathrm{binary} > 0$ for all $k$) and $j$ the loser ($\sigma_{jk}^\mathrm{binary} < 0$ for all $k$), then Eq.~\ref{eq:binary_to_mixed} shows that $\sigma_{ij}^\mathrm{mixed} > 0$, i.e., the winner must always beat the loser in the mixed competition.  Therefore the loser $j$ is guaranteed to go extinct before $i$ can.  But once $j$ goes extinct, the same argument holds for the next-worst strain among the remaining ones, so that it, too, must go extinct before $i$.  Eventually $i$ will be left with just one other strain, in which case $i$ must win because it wins all binary competitions.  Therefore the the winner of the binary competitions inevitably wins the mixed competition.
     
\begin{figure}
\centering\includegraphics[width=\columnwidth]{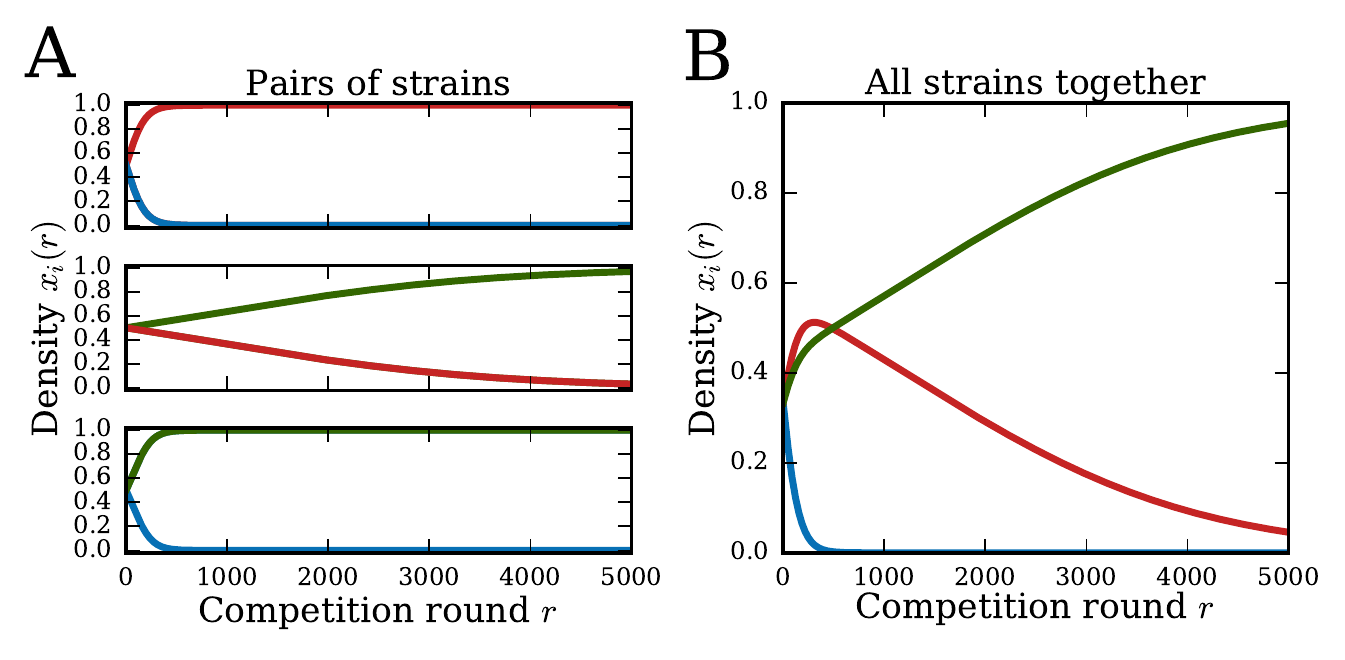}
\caption{
\textbf{Pairwise champion always wins with equal yields.}  
(A)~Density dynamics $x_i(r)$ for binary competitions between three strains (blue, red, green) with a single pairwise champion (green).  
(B)~Density dynamics $x_i(r)$ for a ternary competition starting from equal densities.  
See Sec.~\ref{sec:parameters} for parameter values.
}
\label{fig:pairwise_champion_equal_yields}
\end{figure}

     However, the pairwise champion $i$ may still lose transiently, i.e., $\sigma_{ik}^\mathrm{mixed} < 0$ for some other intermediate strain $k \neq j$.  For example, this occurs in Fig.~\ref{fig:pairwise_champion_equal_yields}, where the green strain beats both blue and red in binary competitions (Fig.~\ref{fig:pairwise_champion_equal_yields}A) but loses transiently to red in the mixed competition at early times (Fig.~\ref{fig:pairwise_champion_equal_yields}B).  This effect is due to a higher-order modification to the selection coefficient from the growth-lag coupling term $s^\mathrm{coupling}_{ijk}$ (Eq.~\eqsformula{}).  However, it only persists until the worst remaining strain (blue) effectively goes extinct, after which green then beats red.
     
\section{Trait constraints for non-transitive competitions}

\begin{figure}
\centering\includegraphics[width=\columnwidth]{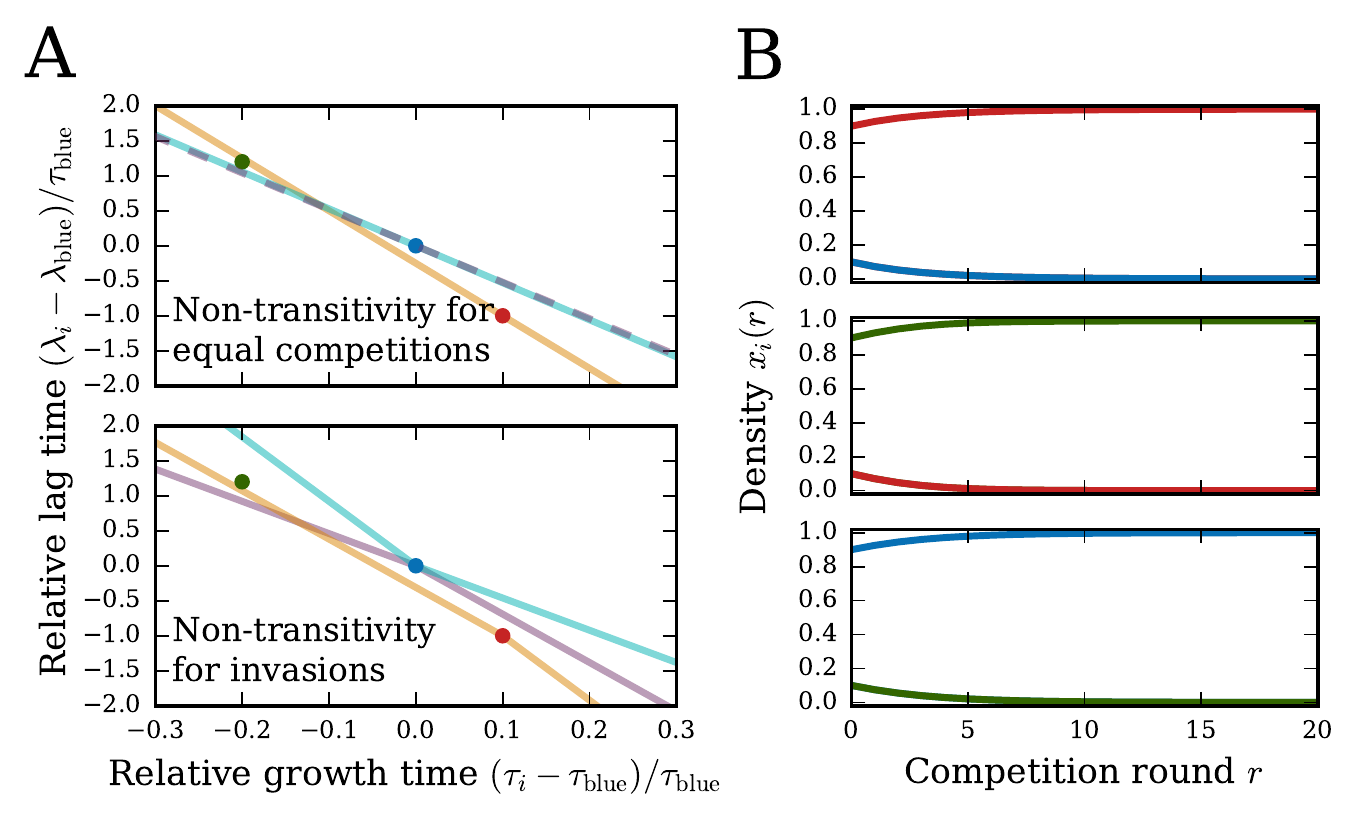}
\caption{
\textbf{Invasions are always transitive.}  
(A)~Diagrams of growth-lag trait space for the same three strains (blue, red, green) as in Fig.~\figbinaryternarycomparisons{}C,D.  The dots mark the same three strains in both top and bottom panels.  For non-transitivity to occur, the red strain must lie below the violet line (so that it beats the blue strain) and the green strain must lie both below the orange line (so that it beats the red strain) and above the cyan line (so that it loses to the blue strain).  The top panel shows these constraints for competitions starting at equal densities (Fig.~\figbinaryternarycomparisons{}C, Eq.~\ref{eq:nontransitivity_equal}), while the bottom panel shows these constraints for invasions (Eq.~\ref{eq:invasion_nontransitivity}).  
(B)~Invasion competitions for each pair of strains from (A).  
See Sec.~\ref{sec:parameters} for parameter values.
}
\label{fig:nontransitivity}
\end{figure}

     Consider a set of three strains: blue, red, and green.  For competitions starting from equal densities to be non-transitive, the binary selection coefficients must satisfy $s_{\red,\blue} > 0$, $s_{\green,\red} > 0$, and $s_{\blue,\green} > 0$, which simplify to
     
\beq
\begin{split}
(\tau_\red - \tau_\blue)\log\left(\rho\Ybar_{\red,\blue}^\mathrm{equal}\right) + (\lambda_\red - \lambda_\blue) & < 0, \\
(\tau_\green - \tau_\red)\log\left(\rho\Ybar_{\green,\red}^\mathrm{equal}\right) + (\lambda_\green - \lambda_\red) & < 0, \\
(\tau_\blue - \tau_\green)\log\left(\rho\Ybar_{\blue,\green}^\mathrm{equal}\right) + (\lambda_\blue - \lambda_\green) & < 0, \\
\end{split}
\label{eq:nontransitivity_equal}
\eeq

\noindent where $\Ybar_{ij}^\mathrm{equal} = \left(Y_i^{-1}/2 + Y_j^{-1}/2\right)^{-1}$ is the harmonic mean of $Y_i$ and $Y_j$ with equal densities (cf. Eq.~\eqtYbar{}).  In the top panel of Fig.~\ref{fig:nontransitivity}A, these three inequalities are represented by the violet, orange, and cyan lines, respectively.

     For invasion competitions to be non-transitive, each strain must beat another strain not just at equal densities, but at all densities.  This results in the following inequalities:
     
\begin{widetext}
\begin{subequations}
\begin{align}
\lambda_\red & < \left\{
\begin{array}{ll}
\lambda_\blue - (\tau_\red - \tau_\blue) \log\left(\rho\min\left[Y_\blue, Y_\red\right]\right) & \text{for } \tau_\red < \tau_\blue \\
\lambda_\blue - (\tau_\red - \tau_\blue) \log\left(\rho\max\left[Y_\blue, Y_\red\right]\right) & \text{for } \tau_\red > \tau_\blue \\
\end{array}
\right. \label{eq:violet}\\
\lambda_\green & < \left\{
\begin{array}{ll}
\lambda_\red - (\tau_\green - \tau_\red) \log\left(\rho\min\left[Y_\red, Y_\green\right]\right) & \text{for } \tau_\green < \tau_\red \\
\lambda_\red - (\tau_\green - \tau_\red) \log\left(\rho\max\left[Y_\red, Y_\green\right]\right) & \text{for } \tau_\green > \tau_\red \\
\end{array}
\right. \label{eq:orange}\\
\lambda_\green & > \left\{
\begin{array}{ll}
\lambda_\blue - (\tau_\green - \tau_\blue) \log\left(\rho\max\left[Y_\blue, Y_\green\right]\right) & \text{for } \tau_\green < \tau_\blue \\
\lambda_\blue - (\tau_\green - \tau_\blue) \log\left(\rho\min\left[Y_\blue, Y_\green\right]\right) & \text{for } \tau_\green > \tau_\blue \\
\end{array}
\right. \label{eq:cyan}
\end{align}
\label{eq:invasion_nontransitivity}
\end{subequations}
\end{widetext}

\noindent These three inequalities define the violet, orange, and cyan lines, respectively, in the bottom panel of Fig.~\ref{fig:nontransitivity}A.  However, the green traits cannot simultaneously satisfy both Eqs.~\ref{eq:orange} and~\ref{eq:cyan}, which we can show by geometrically arguing that the orange and cyan lines can never intersect.  Without loss of generality we assume the blue strain has the smallest yield ($Y_\blue < Y_\red,Y_\green$).  Now first consider the case where $Y_\red < Y_\green$.  Then the left branch of the orange line has (negative) slope $\log\nu_\red$, while the left branch of the cyan line has the steeper slope $\log\left(\rho Y_\green\right)$.  Thus the lines diverge in this direction.  They also diverge to the right, since the right branch of the orange line has slope $\log\left(\rho Y_\green\right)$, which is steeper than the cyan line's slope of $\log\left(\rho Y_\blue\right)$.  Since the orange line is also constrained to be below the violet line at $\tau = 0$ (by the constraints on the red strain, Eq.~\ref{eq:violet}), the orange and cyan lines therefore never intersect.  A similar argument holds if we flip the ordering of the red and green yields, so that $Y_\green < Y_\red$.  Therefore it is not possible for three strains to invade each other non-transitively.  Figure~\ref{fig:nontransitivity} shows the invasion competitions for the same three strains that are non-transitive in equal competitions (Fig.~\figbinaryternarycomparisons{}C).

\section{Additional parameter values for figures}
\label{sec:parameters}

     \textbf{Figure~\figcoexistence{}.}  Growth and lag times are shown in panel (A).  Yields are $Y_\blue = 500$, $Y_\red = 600$, $Y_\green = 750$, and $Y_\orange = 1000$.  In panel (B), the three values of $\rho$ are 0.75, 1.5, and 2.25.  In panel (C), $\rho = 1.32$.

     \textbf{Figure~\figbinaryternarycomparisons{}.}  (A, B)~Growth times are $\tau_\blue = 1$, $\tau_\red = 0.978$, and $\tau_\green = 1.025$; lag times are $\lambda_\blue = 0.15$, $\lambda_\red = 0.28$, and $\lambda_\green = 0$; and yields are $\rho Y_\blue = \rho Y_\red = 10^3$ and $\rho Y_\green = 200$.  (C, D)~Growth times are $\tau_\blue = 1$, $\tau_\red = 1.1$, and $\tau_\green = 0.8$; lag times are $\lambda_\blue = 1$, $\lambda_\red = 0$, and $\lambda_\green = 2.2$; yields are $\rho Y_\blue = 10^2$, $\rho Y_\red = 10^3$, and $\rho Y_\green = 10^4$.

     \textbf{Figure~\ref{fig:coexistence_freq_space}.}  The growth-lag tradeoff is $c = \log 1000$, and the yields are $Y_1 = 500$, $Y_2 = 600$, $Y_3 = 750$, and $Y_4 = 1000$.

     \textbf{Figure~\ref{fig:stability}:} (A)~Growth times are $\tau_\blue = 1$, $\tau_\red = 1.01$, and $\tau_\green = 1.02$.  (B)~Growth times are $\tau_\blue = 1.02$, $\tau_\red = 1.01$, and $\tau_\green = 1$.  (C)~Growth times are $\tau_\blue = 1.01$, $\tau_\red = 1.02$, and $\tau_\green = 1$.  In all panels the growth-lag tradeoff (which defines the lag times via Eq.~\eqgrowthlagtradeoff{}) is $c = \log(2^{1/4} \times 10^3)$, and the yields are $\rho Y_\blue = 10^3$, $\rho Y_\red = 2^{1/2} \times 10^3$, and $\rho Y_\green = 2\times 10^3$.

     \textbf{Figure~\ref{fig:pairwise_champion_equal_yields}:} Growth times are $\tau_\blue = 1$, $\tau_\red = 1.01$, and $\tau_\green = 2$; lag times are $\lambda_\blue = 6.9195$, $\lambda_\red = 6.8395$, $\lambda_\green = 0$; yields are $\rho Y_\blue = \rho Y_\red = \rho Y_\green = 10^3$.  


\bibliographystyle{unsrt}
\bibliography{References}
